%% file: HIG-18-006_temp.tex
\pdfoutput=1

\documentclass[11pt,twoside,a4paper,cmspaper,final,collab]{cms-tdr}
\def\svnVersion{9f1337f-D}\def\svnDate{2019/10/12}\def\cmsCernNoTag{CERN-EP-2019-105}\def\cmsCernDate{\today}\def\cmsMessage{"Published in Physics Letters B as \href{http://dx.doi.org/10.1016/j.physletb.2019.135087}{\doi{10.1016/j.physletb.2019.135087}.}"}

\begin{document}\cmsNoteHeader{HIG-18-006}

\hyphenation{had-ron-i-za-tion}
\hyphenation{cal-or-i-me-ter}
\hyphenation{de-vices}
\RCS$Revision$
\RCS$HeadURL$
\RCS$Id$
\newlength\cmsFigWidth
\ifthenelse{\boolean{cms@external}}{\setlength\cmsFigWidth{0.99\columnwidth}}{\setlength\cmsFigWidth{0.6\textwidth}}
\ifthenelse{\boolean{cms@external}}{\providecommand{\cmsLeft}{upper}}{\providecommand{\cmsLeft}{left}}
\ifthenelse{\boolean{cms@external}}{\providecommand{\cmsRight}{lower}}{\providecommand{\cmsRight}{right}}

\newcommand{\PaO}{\ensuremath{\Pa_1}\xspace}
\newcommand{\sigmaBR}{\ensuremath{\sigma \mathcal{B}}}
\newcommand{\PgPg}{\ensuremath{{\Pg\Pg}\mathrm{F}}\xspace}

 \newlength\cmsTabSkip\setlength{\cmsTabSkip}{1ex}

\cmsNoteHeader{HIG-18-006}
\title{Search for light pseudoscalar boson pairs produced from decays of
the 125\GeV Higgs boson in final states with two muons and two nearby tracks in $\Pp\Pp$ collisions
at \texorpdfstring{$\sqrt{s}=13\TeV$}{13 TeV}}

\date{\today}

\abstract{
A search is presented for pairs of light pseudoscalar bosons, in the mass range from 4 to 15\GeV,
produced from decays of the 125\GeV Higgs boson. The decay modes considered are final states that
arise when one of the pseudoscalars decays to a pair of tau leptons, and the other one either into
a pair of tau leptons or muons. The search is based on proton-proton collisions collected by
the CMS experiment in 2016 at a center-of-mass energy of 13\TeV that correspond
to an integrated luminosity of 35.9\fbinv.  The $2\mu2\tau$ and $4\tau$ channels are used
in combination to constrain the product of the Higgs boson production cross section and the branching fraction
into $4\tau$ final state, $\sigmaBR$, exploiting the linear dependence of the fermionic coupling strength
of pseudoscalar bosons on the fermion mass. No significant excess is observed
beyond the expectation from the standard model. The observed and expected upper limits
at $95\%$ confidence level on $\sigmaBR$, relative to the standard model
Higgs boson production cross section, are set respectively between 0.022 and 0.23
and between 0.027 and 0.19 in the mass range probed by the analysis.
}

\hypersetup{%
pdfauthor={CMS Collaboration},%
pdftitle={Search for light pseudoscalar boson pairs produced from decays of
 the 125 GeV Higgs boson in final states with two muons and two nearby tracks in pp collisions at sqrt(s)=13 TeV},%
pdfsubject={CMS},%
pdfkeywords={CMS, physics, Higgs boson, NMSSM, 2HD+1S}}

\maketitle

\section{Introduction}
\label{sec:intro}

After the discovery of the 125\GeV Higgs boson ($\PH$)~\cite{Aad:2012tfa,Chatrchyan:2012xdj},
searches for additional Higgs bosons, based on predictions beyond the standard model (SM),
constitute an important part of the scientific program at the CERN Large Hadron Collider (LHC).
The present analysis examines theoretical models that contain two Higgs doublets
and an additional complex singlet Higgs field (denoted hereafter as 2HD+1S), that does not couple
at tree level to fermions or gauge bosons and interacts only with itself and the
Higgs doublets~\cite{MSSM:1,Kaul:1981hi,Barbieri1982343,Nilles1983346,Frere198311,Derendinger1984307,Ellwanger:2009dp,Maniatis:2009re}.
In CP conserving models, which are considered in this Letter, the Higgs sector features seven
physical states, namely three CP-even, two CP-odd, and two charged bosons, where one of the CP-even states corresponds to the H.
This kind of Higgs sector is realized, for example, in next-to-minimal
supersymmetric models that solve the so-called $\mu$ problem of
the minimal supersymmetric extension of the SM~\cite{Kim:1983dt}.
A large set of the 2HD+1S models is allowed by
measurements and constraints set by searches
for additional Higgs bosons and supersymmetric particles~\cite{Belanger:2012tt,Belanger:2012sd,Gunion:2012he,Gunion:2012gc,King:2012is,King:2012tr}.

This Letter addresses specific 2HD+1S models
in which the lightest pseudoscalar boson (\PaO) with mass $2m_{\PaO}<125$\GeV has a large singlet component,
and therefore its couplings to SM particles are significantly reduced.
For this reason, analyses using direct production modes of \PaO, such as
gluon-gluon fusion (\PgPg) or {\cPqb} quark associated production, have limited sensitivity.
The \PaO boson is nonetheless potentially accessible
in the $\PH$ decay to two pseudoscalar bosons.
The \PaO states can be identified via their decay into a pair of
fermions~\cite{Ellwanger:2005uu,Ellwanger:2003jt,Ellwanger:2004gz,Belyaev:2008gj,Belyaev:2010ka,aDecays,Almarashi:2012ri,Almarashi:2011qq}.
Constraints on the $\PH$  couplings allow a branching fraction for $\PH$ decays into non-SM particles as large
as 34\%~\cite{Khachatryan:2016vau}, which can potentially accommodate the $\PH\to \PaO\PaO$
decay at a rate sufficiently high for detection at the LHC.

Several searches for $\PH\to \PaO \PaO$ decays have been performed
in the ATLAS and CMS experiments in Run 1 (8\TeV) and Run 2 (13\TeV) of LHC,
exploiting various decay modes of the \PaO boson, and probing different ranges of its
mass~\cite{Khachatryan:2017mnf,Khachatryan:2015nba,Khachatryan:2015wka,Sirunyan:2018mbx,Sirunyan:2018pzn,Aaboud:2016oyb,Aad:2015sva,Aad:2015bua,Aad:2015oqa,Aaboud:2018gmx,Aaboud:2018fvk,Sirunyan:2018mot,Sirunyan:2018mgs,Aaboud:2018esj}.
These searches found no significant deviation from the expectation
of the SM background and upper limits were set on the product of  the production cross section and the branching fraction
for signal resulting in constraints on parameters of the 2HD+1S models.

This analysis presents a search for light \PaO bosons in the decay channels
$\PH\to\PaO\PaO\to 4\tau/2\mu 2\tau$, using
 data corresponding to an integrated luminosity of 35.9\fbinv, collected with the
CMS detector in 2016 at a center-of-mass energy of 13\TeV. The analysis covers the mass range from 4 to 15\GeV
and employs a special analysis strategy to select and
identify highly Lorentz-boosted muon or tau lepton pairs with overlapping decay products.
The study updates a similar one performed by the CMS Collaboration in Run 1~\cite{Khachatryan:2015nba}, and complements other recent CMS searches for the $\PH\to\PaO\PaO$ decay performed in Run 2 data
in the $2\mu 2\tau$~\cite{Sirunyan:2018mbx}, $2\tau 2{\cPqb}$~\cite{Sirunyan:2018pzn}, $2\mu 2{\cPqb}$~\cite{Sirunyan:2018mot} and $4\mu$~\cite{Sirunyan:2018mgs} final states, covering
respective mass ranges of $0.25 <m_{\PaO}<3.40\GeV$ for the $4\mu$ final state and $15.0<m_{\PaO}<62.5\GeV$ for the $2\mu 2\tau,2\tau 2{\cPqb}$, and $2\mu 2{\cPqb}$
final states.

The branching fraction $\PaO\to \tau\tau$ depends on the details of the model, namely the parameter $\tan\beta$, the ratio of vacuum expectation values of the two Higgs doublets, and on which Higgs doublet couples to either charged leptons, up-type quarks or down-type quarks~\cite{Curtin:2013fra}.
In Type-II 2HD+1S models, where one Higgs doublet couples to up-type fermions while the other couples to down-type fermions,
the $\PaO\to \tau\tau$ decay rate gets enhanced at large values of $\tan\beta$. The branching fraction of this decay reaches values
above 90\% at $\tan\beta>3$ for $2m_{\tau} < m_{\PaO} <2m_{\cPqb}$, where $m_\tau$ is the mass of the
tau lepton and $m_{\cPqb}$ is the mass of the bottom quark.
For higher values of $m_{\PaO}$ the branching fraction decreases to 5--6\%
since the decay into a pair of bottom quarks becomes kinematically possible and overwhelms the decay into a pair of
tau leptons. However, in some of the 2HD+1S models the $\PaO\to \tau\tau$ decay may be dominant even above the
$\PaO\to {\cPqb}{\cPaqb}$ decay threshold. This is realized, e.g., for
$\tan\beta>1$ in the Type-III 2HD+1S models, where one Higgs doublet couples to charged leptons, whereas the other doublet
couples to quarks~\cite{Curtin:2013fra}.

\begin{figure*}[htb]{
\begin{center}{
\includegraphics[width=0.6\textwidth]{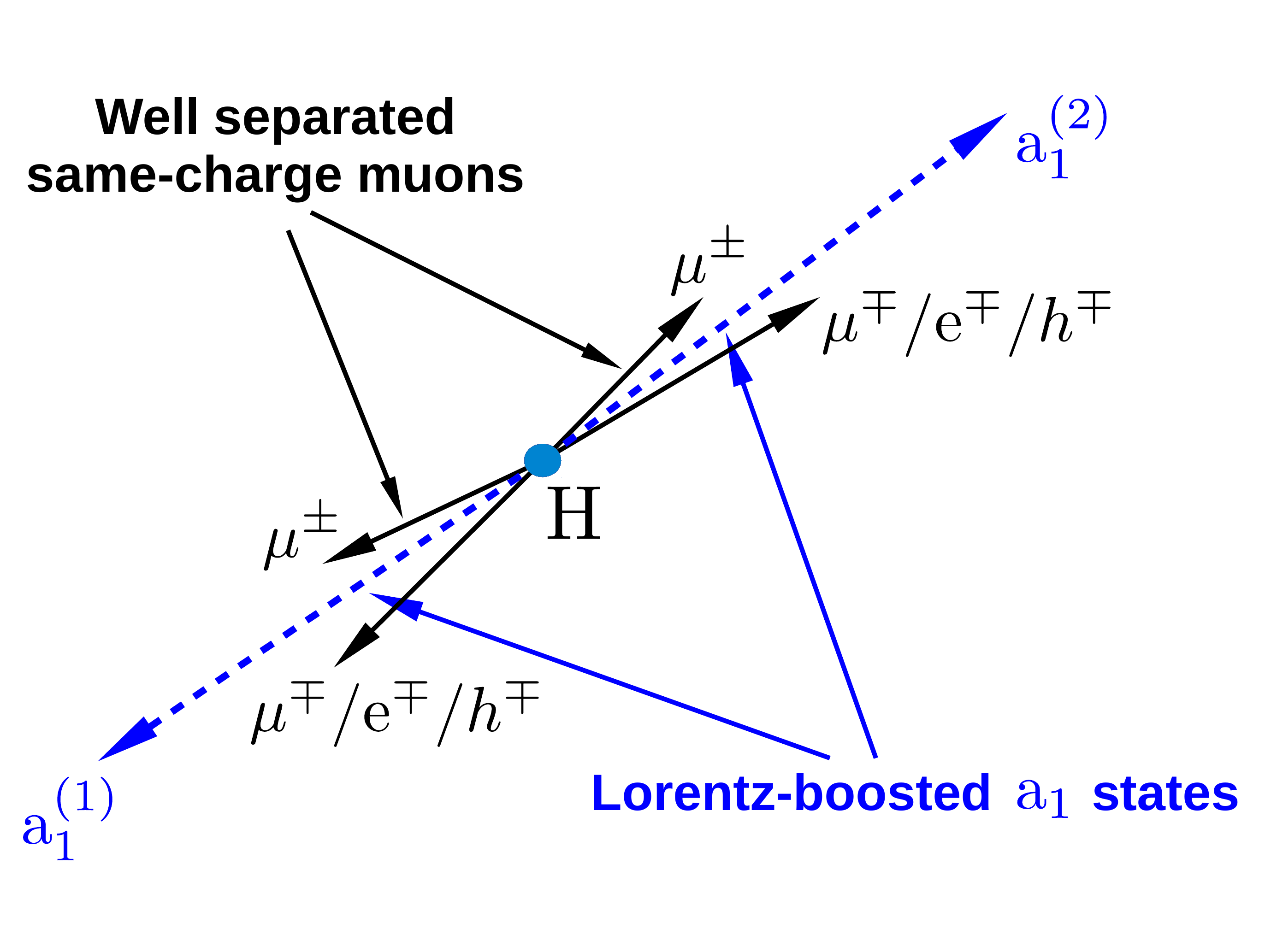}
\caption{
Illustration of the signal topology, in which the $\PH$  decays into two \PaO bosons, where
one \PaO boson decays into a pair of tau leptons, while the other one decays into a pair of muons or a pair of tau leptons.
The analyzed final state consists of one muon and an oppositely charged track in each \PaO decay.
\label{fig:topology}
}
}
\end{center}
}
\end{figure*}

The signal topology targeted by the present analysis is illustrated in Fig.~\ref{fig:topology}.
Each \PaO boson is identified by
the presence of a muon and only one additional charged particle, the objective of this approach being the decay channels
$\PaO\to\mu\mu$ and $\PaO\to\tau_{\mu}\tau_{\text{one-prong}}$.
The $\tau_\mu$ denotes the muonic tau lepton decay, and
$\tau_{\text{one-prong}}$ stands for its leptonic or one-prong hadronic decay. The three-prong modes are not 
used because of the very high QCD multijet background and lower reconstruction signal efficiency.

Given the large difference in mass between the \PaO and the
$\PH$ states, the \PaO bosons will be produced highly Lorentz-boosted, and
their decay products are highly collimated. This will result in a signature with two muons, each of which
is accompanied by a nearby particle of opposite charge. The search focuses primarily
on the dominant \PgPg process, in which the
$\PH$ state is produced with relatively small transverse
momentum \pt, and the \PaO pseudoscalars are emitted nearly
back-to-back in the transverse plane, with
a large separation in azimuth $\phi$
between the particles originating from one of the \PaO decays and those of the other \PaO.
In the \PgPg process, the \PH can be also produced with a relatively high Lorentz boost when a hard
gluon is radiated from the initial-state gluons or from the heavy-quark loop. In this case,
the separation in $\phi$ is reduced, but the separation in pseudorapidity $\eta$ can be large.
The analysis therefore searches for a signal in a
sample of same-charge (SC) dimuon events with large angular separation between the muons, where
each muon is accompanied by one nearby oppositely charged particle originating from the same \PaO decay.
The requirement of having SC muons in the event largely suppresses background from the
top-quark-pair, Drell--Yan, and diboson production.
This requirement also facilitates the implementation of a dedicated SC dimuon trigger with relatively low thresholds and acceptable rates
as described in Section~\ref{Sec:Selection}.

\section{CMS detector}
\label{sec:detector}

The central feature of the CMS detector is a superconducting solenoid
of 6\unit{m} internal diameter, providing a magnetic field of 3.8\unit{T}.
Within the solenoid volume are a silicon pixel and strip tracker, a
lead tungstate crystal electromagnetic calorimeter, and a brass and scintillator
hadron calorimeter, each composed of a barrel and two endcap sections.
Forward calorimeters extend the $\eta$ coverage provided by the
barrel and endcap detectors. Muons are detected in gas-ionization chambers
embedded in the steel flux-return yoke outside the solenoid.

Events of interest are selected using a two-tiered trigger system~\cite{Khachatryan:2016bia}.
The first level, composed of custom hardware processors, uses information from the
calorimeters and muon detectors to select events at a rate of around 100\unit{kHz} within a
time interval of less than 4\mus. The second level, known as the high-level trigger,
consists of a farm of processors running a version of the full event reconstruction software optimized for fast processing, and reduces the event rate below 1\unit{kHz} before data storage.

A more detailed description of the CMS detector, together with a definition of the coordinate system used and the relevant kinematic variables, can be found in Ref.~\cite{Chatrchyan:2008zzk}.

\section{Simulated samples}
\label{sec:data set}

For the simulation of the dominant \PgPg production process,
the Monte Carlo (MC) event generators \PYTHIA (v.8.212)~\cite{Sjostrand:2014zea} and \MGvATNLO (v.2.2.2)~\cite{Alwall:2014hca} are
used in order to model the $\PH \to \PaO \PaO \to 4\tau$ and $\PH \to \PaO \PaO \to 2\mu 2\tau$ signal events, respectively.
For both decay modes the \pt distribution of the $\PH$  emerging from \PgPg
is reweighted with next-to-next-to-leading order (NNLO) $K$ factors obtained by the program \textsc{HqT} (v2.0)~\cite{Bozzi:2005wk,deFlorian:2011xf}
with NNLO NNPDF3.0 parton distribution functions (PDF)~\cite{Ball:2014uwa},
hereby taking into account the more precise spectrum calculated to NNLO with resummation to next-to-next-to-leading-logarithms order.
Subdominant contributions from other production modes of $\PH$, namely vector boson fusion process (VBF), vector boson associated
production (VH) and top quark pair associated production ({$\ttbar\PH$}) are estimated using the \PYTHIA (v.8.212) generator.

The backgrounds from diboson production and quantum chromodynamics production of multijet (QCD multijet)
are simulated with the \PYTHIA (v.8.212) generator. Inclusive \PZ and \PW boson
production processes are generated with \MGvATNLO  (v.2.2.2).
The single-top and \ttbar production are generated at Next-to-LO (NLO) with the \POWHEG (v.2.0) generator ~\cite{Nason:2004rx,Frixione:2007vw,Alioli:2010xd,Re:2010bp,Alioli:2009je}.
The set of PDF used is NLO NNPDF3.0 for NLO samples, and LO NNPDF3.0 for LO samples~\cite{Ball:2014uwa}.

Showering and hadronization are carried out by the \PYTHIA  (v.8.212) generator
with the CUETP8M1 underlying event tune~\cite{Khachatryan:2015pea},
while a detailed simulation of the CMS detector is based on the \GEANTfour~\cite{geant4} package.

\section{Event selection}
\label{Sec:Selection}

Events are selected using a SC dimuon trigger with \pt thresholds of 17 (8)\GeV for the leading (subleading) muon.
To pass the high-level trigger, the tracks of the two muons are additionally required
to have points of closest approach to the beam axis within 2\mm of each other
along the longitudinal direction.

Events are reconstructed with the particle-flow (PF) algorithm~\cite{Sirunyan:2017ulk}
which aims to identify and reconstruct individual particles as photons,
charged hadrons, neutral hadrons, electrons, or muons (PF objects).
The proton-proton ($\Pp\Pp$) interaction vertices are reconstructed
using a Kalman filtering technique~\cite{Adam:934067,Chatrchyan:2014fea}.
Typically more than one such vertex is reconstructed because of
multiple $\Pp\Pp$ collisions within the same or neighbouring bunch crossings.
The mean number of such interactions per bunch crossing was 23 in 2016.

The reconstructed vertex with the largest value of summed physics-object $\pt^2$ is taken to be the primary interaction vertex (PV).
The physics objects are the jets, clustered using
the jet-finding algorithm~\cite{Cacciari:2008gp,Cacciari:2011ma} with the tracks assigned
to the vertex as inputs, and the associated missing transverse momentum,
taken as the negative vector sum of the \pt of those jets.
Events must contain at least two SC muons reconstructed with the PF algorithm, which have to fulfil the following requirements.

\begin{itemize}
\item{The pseudorapidity of the leading (higher \pt) and the subleading (lower \pt) muons must be $\abs{\eta} < 2.4$}.

\item{The \pt of the leading (subleading) muon must exceed 18 (10)\GeV}.

\item{The transverse (longitudinal) impact parameters of muons with respect to the PV are required to be $\abs{d_0}<0.05$ ($\abs{d_z}<0.1$)\cm}.

\item{The angular separation between the muons is
$\Delta{\mathrm{R}}=\sqrt{\smash[b]{ (\Delta\phi)^2+(\Delta\eta^2)}}>2$.}
\end{itemize}

If more than one SC muon pair is found in the event to satisfy these requirements,
the pair with the largest scalar sum of muon \pt  is chosen.

In the next step, the analysis employs information about tracks associated with the reconstructed charged PF objects, excluding
the pair of SC muons.
Selected muons and tracks are used to build and isolate
candidates for the $\PaO\to \tau_\mu\tau_{\text{one-prong}}$
or $\PaO\to\mu\mu$ decays (referred to as \PaO candidates
throughout the Letter). Three types of tracks are considered in the analysis.

\begin{itemize}
\item{``Isolation" tracks are used to define isolation requirements imposed on \PaO candidates and have to fulfil the following criteria:
$\pt>1$\GeV, $\abs{\eta}<2.4$, $\abs{d_0}<1$\cm, $\abs{d_z}<1$\cm.}
\item{``Signal" tracks are selected among ``isolation" tracks to build \PaO candidates.
These tracks must have $\pt>2.5$\GeV, $\abs{\eta}<2.4$, $\abs{d_0}<0.02$\cm, $\abs{d_z}<0.04$\cm.}
\item{``Soft" tracks are also a subset of ``isolation" tracks. They are
utilized to define one of the
sideband regions, used for the construction of the background model, as described in Section~\ref{Sec:bkdg_corr}. ``Soft"
tracks must satisfy the requirements: $1.0<\pt<2.5$\GeV, $\abs{\eta}<2.4$, $\abs{d_0}<1$\cm, $\abs{d_z}<1$\cm.}
\end{itemize}

A track is regarded as being nearby a muon if the angular separation $\Delta{\mathrm{R}}$ between them is smaller than 0.5.
Each muon of the SC pair is required to have one nearby ``signal" track with a charge opposite to its charge.
This muon-track system is accepted as an \PaO candidate if no additional ``isolation" tracks are found
in the $\Delta{\mathrm{R}}$ cone of 0.5 around
the muon momentum direction. The event is selected in the final sample if it contains two \PaO candidates. The set of selection
requirements outlined above defines the signal region (SR).

The expected signal acceptance and signal yield for a few representative values of $m_{\PaO}$ are reported in
Table~\ref{tab:signal_selection}. The signal yields are computed
for a benchmark value of the branching fraction, ${\mathcal{B}}(\PH \to \PaO \PaO)  {\mathcal{B}}^{2} (\PaO \to \tau \tau)=0.2$ and assuming that
the $\PH$ production cross section is the one predicted in the SM. Contributions from the \PgPg, VBF, VH and
{$\ttbar\PH$} processes are summed up. The yield of the $2\mu 2\tau$ signal is estimated
under the assumption that the partial widths of the $\PaO\to\mu\mu$ and $\PaO\to\tau\tau$ decays
satisfy the relation~\cite{aDecays}
\begin{equation}
\frac{\Gamma(\PaO \to \mu\mu)}{\Gamma(\PaO \to \tau\tau)}
=\frac{m^2_{\mu}}{m^2_{\tau}\sqrt{1-\big(2m_{\tau}/m_{\mathrm{a_{1}}}\big)^2}}.
\label{eq:width_ratios}
\end{equation}
The ratio of branching fractions of the $\PaO\PaO \to 2\mu 2\tau$ and $\PaO\PaO \to 4\tau$
decays is computed through the ratio of the partial widths $\Gamma(\PaO \to \mu\mu)$ and
$\Gamma(\PaO \to \tau\tau)$ as
\begin{equation}
\frac{{\mathcal{B}}(\PaO\PaO \to 2\mu 2\tau)}{{\mathcal{B}}(\PaO\PaO \to 4\tau)}
= 2\frac{{\mathcal{B}}(\PaO \to \mu\mu)}{{\mathcal{B}}(\PaO \to \tau\tau)}
= 2\frac{\Gamma(\PaO \to \mu\mu)}{\Gamma(\PaO \to \tau\tau)}.
\label{eq:br_ratios}
\end{equation}
The factor of 2 in Eq.~(\ref{eq:br_ratios}) arises from two possible decays, $\PaO^{(1)}\PaO^{(2)} \to 2\mu 2\tau$ and
$\PaO^{(1)}\PaO^{(2)} \to 2\tau 2\mu$, that produce the final state with two muons and two tau leptons.
The ratio in Eq.~(\ref{eq:br_ratios}) ranges from about 0.0073 at $m_{\PaO}=15\GeV$ to 0.0155 at $m_{\PaO}=4\GeV$.

The contribution from the  $\PH\to  \PaO\PaO \to 4\mu$
decay is estimated taking into account Eq.~(\ref{eq:width_ratios}).
It ranges between 0.4 and 2\% of the total signal yield in the $2\mu 2\tau$ and $4\tau$ final states, depending on the probed
mass of the \PaO boson. This contribution is not considered in the present analysis.

The number of observed events selected in the SR amounts to 2035.
A simulation-based study shows that the QCD multijet events dominate the sample of events selected in the SR.
Contribution from other background sources constitutes about $1\%$ of
events selected in the SR.

\begin{table*}[htb]
\topcaption{
The signal acceptance and the number of expected signal events after selection in the SR. The
number of expected signal events is computed for a benchmark value of branching fraction,
${\mathcal{B}}(\PH \to \PaO \PaO)  {\mathcal{B}}^{2} (\PaO \to \tau \tau)=0.2$
and assuming that the $\PH$ production cross section is the one predicted in the SM.
The quoted uncertainties for predictions from simulation include only statistical ones.}
\label{tab:signal_selection}

\begin{center}
\begin{tabular}{ccccc}
  & \multicolumn{2}{c}{Acceptance $\times 10^{4}$}&\multicolumn{2}{c}{Number of events}  \\
$m_{\mathrm{a}_{1}}$ [\GeV] & $4\tau$ & $2\mu 2\tau$ & $4\tau$ & $2\mu 2\tau$ \\
\hline
$\phantom{1}$4 & 3.29 $\pm$ 0.16 &            89.3 $\pm$ 1.4 & 129.9            $\pm$ 6.2 & 54.7           $\pm$ 0.9 \\
$\phantom{1}$7 & 2.50 $\pm$ 0.14 &            69.0 $\pm$ 1.4 & \phantom{1}98.8  $\pm$ 5.5 & 22.5           $\pm$ 0.5 \\
            10 & 1.46 $\pm$ 0.11 &            47.1 $\pm$ 1.2 & \phantom{1}57.8  $\pm$ 4.2 & 14.2           $\pm$ 0.4 \\
            15 & 0.21 $\pm$ 0.04 &  \phantom{8}3.5 $\pm$ 0.3 & \phantom{11}8.5  $\pm$ 1.1 & \phantom{1}1.0 $\pm$ 0.1 \\
\end{tabular}
\end{center}
\end{table*}

\begin{figure}[hbtp]
\begin{center}
\includegraphics[width=\cmsFigWidth]{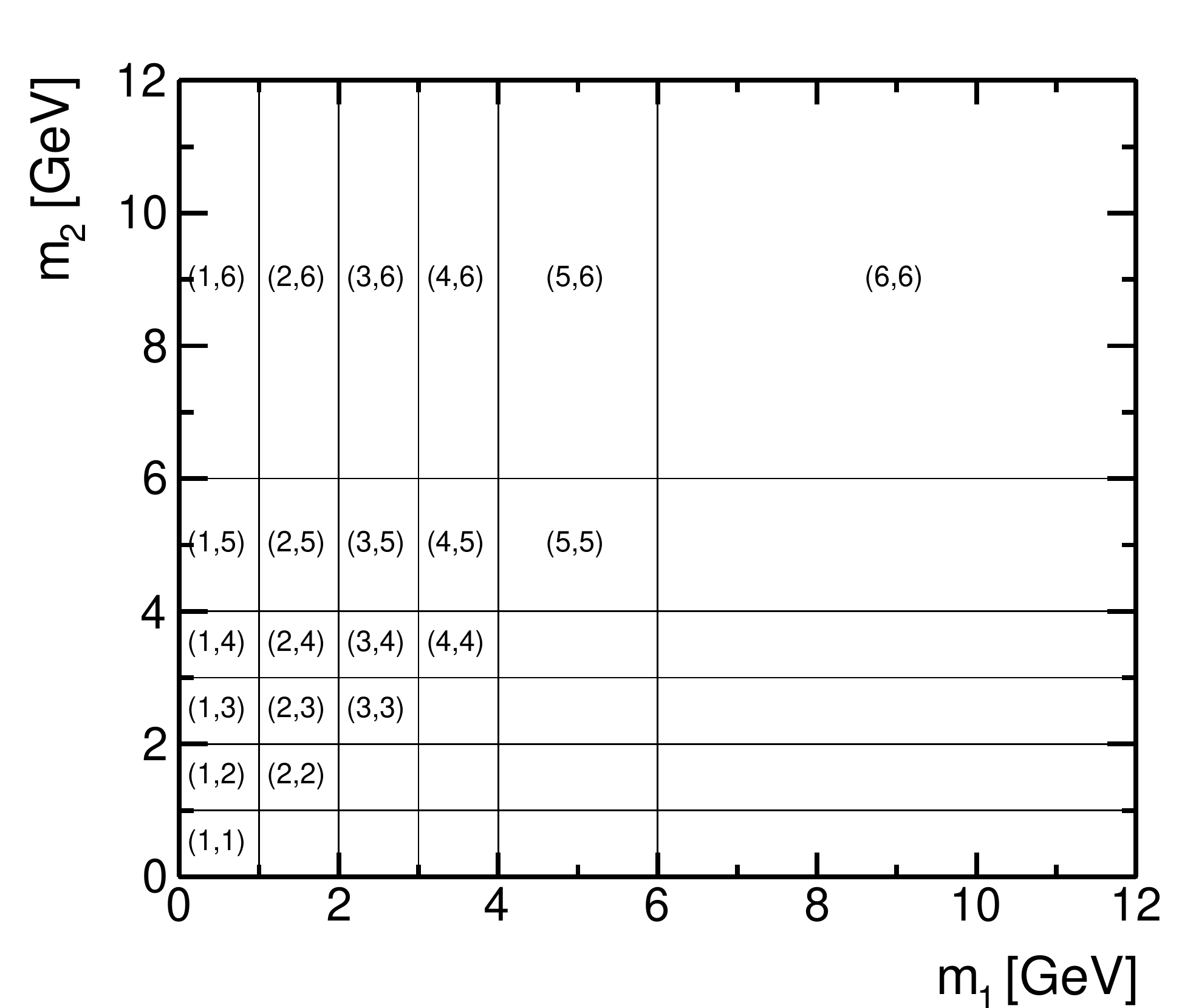}
\caption{Binning of the 2D ($m_1,m_2$) distribution.}
\label{fig:binning}
\end{center}
\end{figure}

The two-dimensional (2D) distribution of the invariant masses of the muon-track systems, constituting \PaO candidates,
is used to discriminate between signal and the dominant QCD multijet background in the signal extraction procedure.
The 2D distribution is filled with a pair
of the muon-track invariant masses ($m_1,m_2$), ordered by their value, $m_2>m_1$. The binning of the 2D
distribution adopted in the analysis is illustrated in Fig.~\ref{fig:binning}. As $m_2$ is required to exceed $m_1$,
only $(i,j)$ bins with $j\ge i$ are filled in the 2D distribution, yielding in total $\mathrm{6 (6+1)/2=21}$ independent bins.
Bins $(i,6)$ with $i=1,5$ contain all events with $m_2>6\GeV$. Bin $(6,6)$ contains all events with $m_{1,2}>6\GeV$.

\section{Modeling background}
\label{Sec:Bkgd}
A simulation-based study reveals that the sample of SC muon pairs selected as described in Section~\ref{Sec:Selection}, but without
requiring the presence of \PaO candidates, is dominated by QCD multijet events, where
about 85\% of all selected events contain bottom quarks in the final state.
The SC muon pairs in these events originate mainly from the following sources:
\begin{itemize}
\item{muonic decay of a bottom hadron in one bottom quark jet and
cascade decay of a bottom hadron into a charm hadron with a subsequent
muonic decay of the charm hadron in the other bottom quark jet;}
\item{muonic decay of a bottom hadron in one bottom quark jet
and decay of a quarkonium state into a pair of muons in the other
jet;}
\item{muonic decay of a bottom hadron in one bottom quark jet
and muonic decay of a $\PBz$ meson in the other bottom quark jet.
The SC muon pair in this case may appear as a result of $\PBz$--$\PaBz$ oscillations.}
\end{itemize}

The normalized 2D ($m_1,m_2$) distribution for the muon-track pairs with $m_2>m_1$ is represented in the sample of background events by a binned template constructed using the following relation
\begin{equation}
\begin{split}
f_\text{2D}(i,j) & = C(i,j) (\,f_\text{1D}(i) f_\text{1D}(j))^{\text{sym}} , \\
(\,f_\text{1D}(i)  f_\text{1D}(i))^{\text{sym}} & = \,f_\text{1D}(i)   f_\text{1D}(i),  \\
 (\,f_\text{1D}(i)  f_\text{1D}(j))^{\text{sym}} & = \,f_\text{1D}(i)   f_\text{1D}(j) + \,f_\text{1D}(j)   f_\text{1D}(i) \\
 &= 2f_\text{1D}(i)f_\text{1D}(j),\qquad \text{if $j>i$},\\
\end{split}
\label{eq:QCDshape}
\end{equation}
where

\begin{itemize}
\item{$f_\text{2D}(i,j)$ is the content of the bin $(i,j)$ in the normalized 2D ($m_1,m_2$) distribution;}
\item{$f_\text{1D}(i)$ is the content of bin $i$ in the normalized one-dimensional (1D) distribution of the muon-track invariant mass;}
\item{$C(i,j)$ is a symmetric matrix, accounting for possible correlation between $m_1$ and $m_2$, the elements of the matrix $C(i,j)$ are referred to as ``correlation factors'' in the following.}
\end{itemize}

The condition $C(i,j)=1$ for all
bins $(i,j)$ would indicate an absence of correlation between $m_1$ and $m_2$.
We sum the contents of the nondiagonal bins $(i,j)$ and $(j,i)$ in the Cartesian product
$f_\text{1D}(i) f_\text{1D}(j)$ to account for the fact that each event enters the 2D ($m_1,m_2$) distribution
with ordered values of the muon-track invariant masses.

By construction the background model estimates the dominant QCD multijet production
as well as small contributions from other processes.

Multiple control regions (CRs) are introduced in order to derive and validate the modeling of $f_\text{1D}(i)$ and $C(i,j)$.
The CRs are defined on the basis of a modified isolation criteria applied to one or both muon-track pairs.
The isolation criteria are specified by the multiplicity of ``isolation" tracks in the cone
of $\Delta\mathrm{R}=0.5$ around the muon momentum direction. The summary of all CRs used to derive and validate the modeling
of background shape is given in Table~\ref{tab:CR}.

\begin{table*}[htb]
\topcaption{Control regions used to construct and validate the background model.
The symbols $N_\text{sig}$, $N_\text{iso}$ and $N_\text{soft}$ denote the
number of ``signal", ``isolation" (which are a subset of ``signal" tracks) and ``soft" tracks, respectively, within
a cone of $\Delta\mathrm{R}=0.5$ around the muon momentum direction. 
The last row defines the SR.}
\label{tab:CR}
\begin{center}
\begin{tabular}{lcccr}
Control region  & First $\mu$ & Second $\mu$  &Purpose & Observed events             \\
\hline

$N_{23}$               & $N_\text{iso} = 1$, $N_\text{sig}=1$    & $N_\text{iso}=2,3$             & Determination of $f_\text{1D}(i)$   & 62\,438 \\
$N_\text{iso,2}=1$   & $N_\text{iso}>1$, $N_\text{sig}\ge 1$ & $N_\text{iso}=1$, $N_\text{sig}=1$ & Validation of $f_\text{1D}(i)$     & 472\,570 \\
$N_\text{iso,2}=2,3$ & $N_\text{iso}>1$, $N_\text{sig}\ge 1$ & $N_\text{iso}=2,3$                   & Validation of $f_\text{1D}(i)$     & 17\,667\,900\\
$N_{45}$               & $N_\text{iso}=1$, $N_\text{sig}=1$    & $N_\text{iso}=4,5$                   & Assessment of                     \\
                       &                                           &                                        & systematics in $f_\text{1D}(i)$    & 52\,437 \\

    \multicolumn{4}{c}{ Both muons  } \\
Loose-Iso            &  \multicolumn{2}{c}{ $N_\text{sig}=1$, $N_\text{soft}=1,2$ }  & Determination of $C(i,j)$ & 35\,824 \\

Signal region        &  \multicolumn{2}{c}{ $N_\text{sig}=1$, $N_\text{iso}=1$ }   & Signal extraction  & 2\,035 \\

\hline
\end{tabular}

\end{center}
\end{table*}

\subsection{Modeling of \texorpdfstring{$f_\text{1D}(i)$}{f[1Di]}}
\label{Sec:bkgd_1D}

The $f_\text{1D}(i)$ distribution is modeled using the $N_{23}$ CR.
Events in this CR pass the SC dimuon selection and contain only one \PaO candidate composed of the
isolated ``signal" track and muon (first muon). The invariant mass of the first muon and associated track enters
the $f_\text{1D}(i)$ distribution.
Another muon (second muon) is required to be accompanied by either two or three nearby ``isolation" tracks.
The simulation shows that more than 95\% of events selected in the CR $N_{23}$ are QCD multijet events, 
while the remaining 5\% is coming from \ttbar, Drell-Yan and other electroweak processes.
The modeling of the $f_\text{1D}(i)$ template is based on the hypothesis that the kinematic distributions for
the muon-track system, making up an \PaO candidate (the first muon and associated track),
are weakly affected by the isolation requirement imposed on the second muon; therefore the $f_\text{1D}(i)$ distribution of
the muon-track system forming an \PaO candidate is expected to be similar in the SR and the $N_{23}$ CR.

\begin{figure}[hbt!]
\begin{center}
\includegraphics[width=0.45\textwidth]{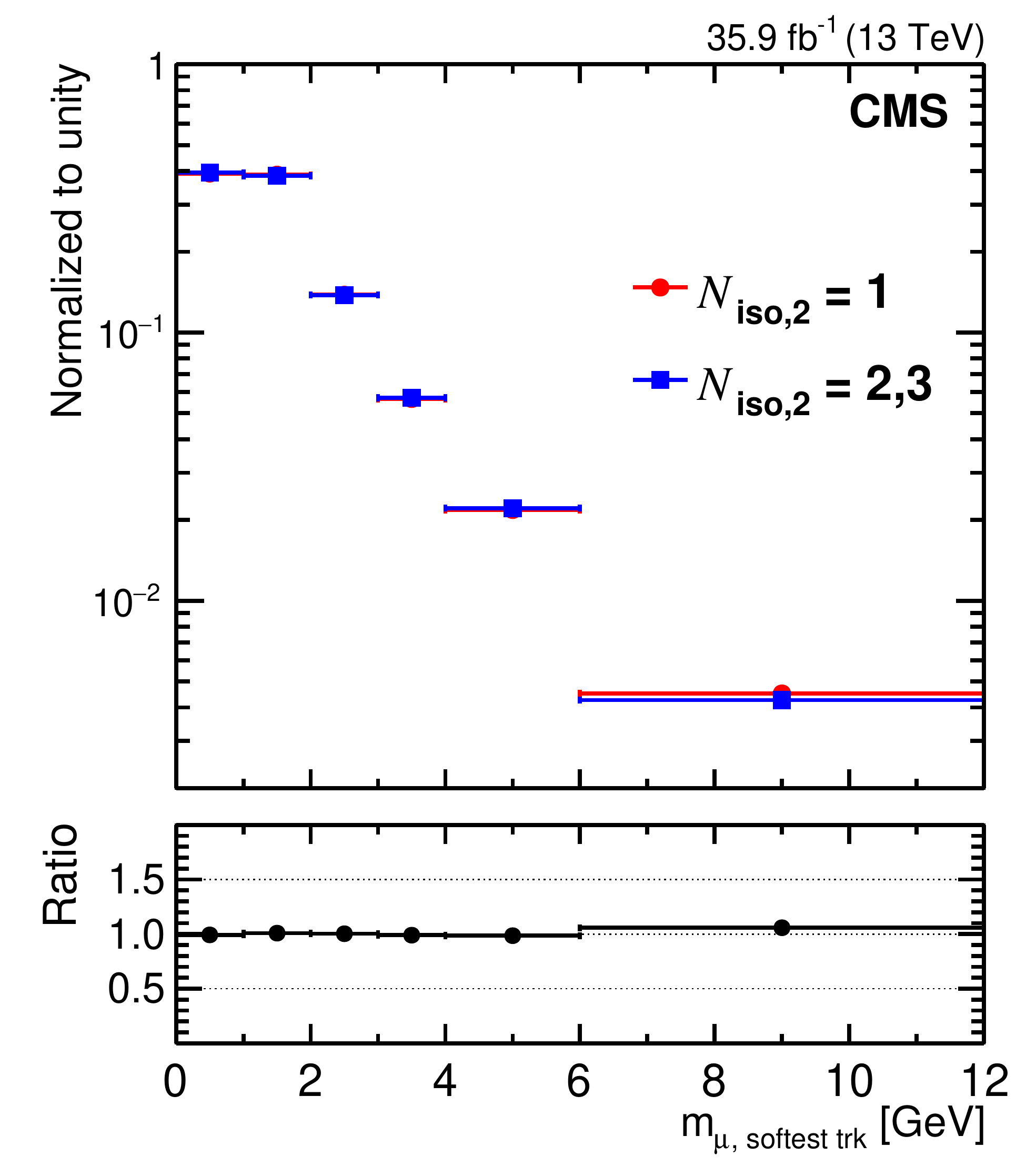}
\includegraphics[width=0.45\textwidth]{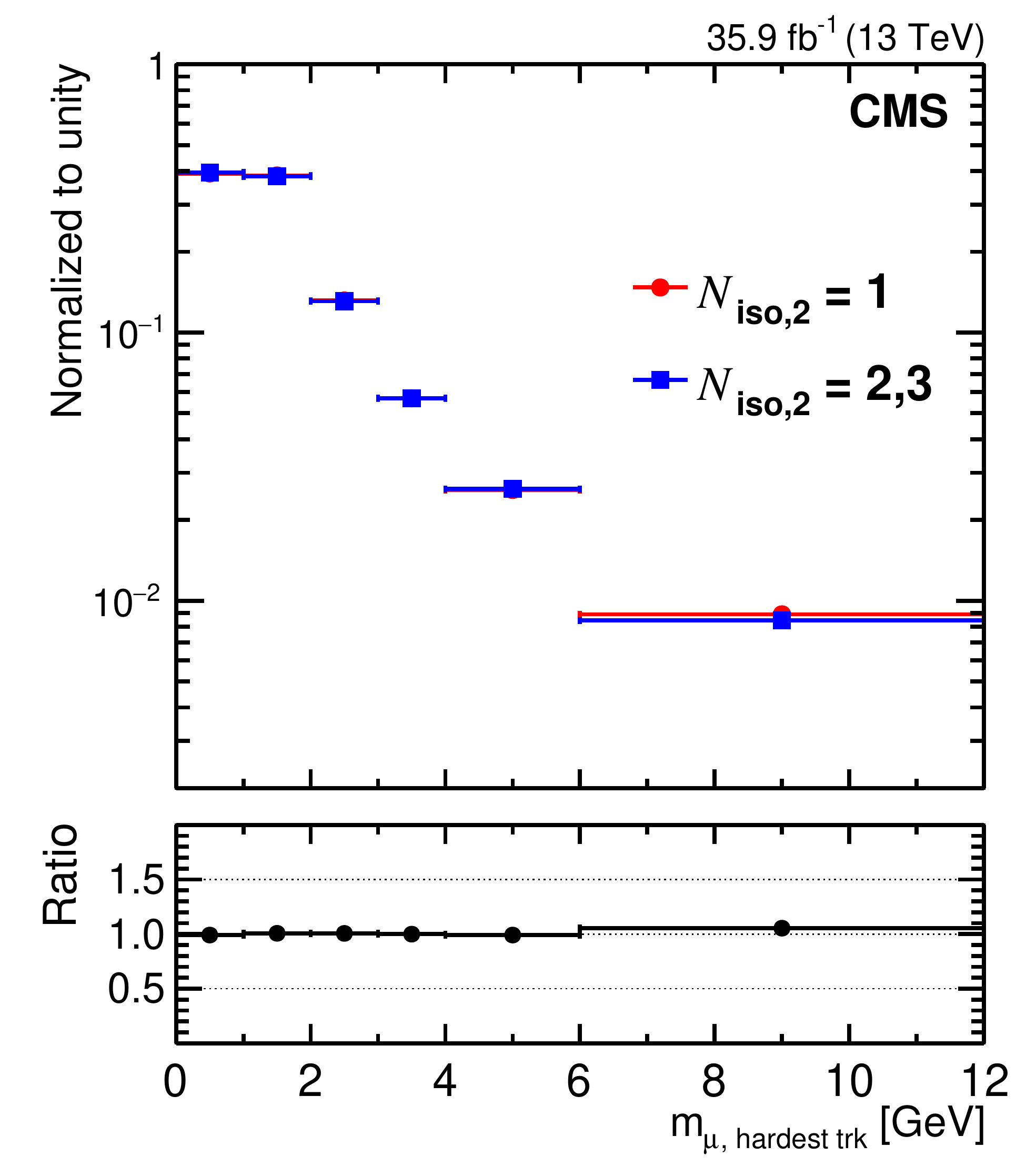}
\caption{
The observed invariant mass distribution, normalized to unity, of the first muon and the softest (\cmsLeft) or
hardest (\cmsRight) accompanying ``signal" track
for different isolation requirements imposed on the second muon:
when the second muon has only one accompanying ``isolation" track ($N_\text{iso,2}=1$; circles);
or when it has two or three accompanying ``isolation" tracks ($N_\text{iso,2}=2,3$; squares).
}
\label{fig:MassShapesIso}
\end{center}
\end{figure}

This hypothesis is verified in control regions labelled $N_\text{iso,2}=1$ and $N_\text{iso,2}=2,3$.
Events are selected in these CR if one of the muons (first muon) has more than one ``isolation" track ($N_\text{iso} > 1$). 
At least one of these ``isolation" tracks should also fulfil the criteria imposed on the ``signal" track.
As more than one of these tracks can pass the criteria imposed on ``signal" tracks,
two scenarios have been investigated, namely using either the lowest or the highest \pt ``signal"  tracks
 (``softest" and ``hardest") to calculate the muon-track invariant mass. If only one ``signal"  track
is found nearby to the first muon, the track is used both as the ``hardest" and the ``softest" signal track.
For the second muon, two isolation requirements are considered: when
the muon is accompanied by only one ``signal" track and the muon-track system
is isolated as in the SR (CR $N_\text{iso,2} = 1$), or when it is accompanied by
two or three ``isolation" tracks as in the CR $N_{23}$ (CR $N_\text{iso,2} = 2, 3$). The invariant mass
distributions of the first muon and the softest or hardest accompanying track are then
compared for the two different isolation requirements on the second muon, $N_\text{iso,2} = 1$
and $N_\text{iso,2} = 2, 3$. The results of this study are illustrated in Fig.~\ref{fig:MassShapesIso}.
In both cases, the invariant mass distributions differ in each
bin by less than 6\%. This observation indicates that the invariant mass of the muon-track system,
making up an \PaO candidate, weakly depends on the isolation requirement imposed on the
second muon, thus supporting the assumption that the $f_\text{1D}(i)$ distribution can be
determined from the $N_{23}$ CR.

\begin{figure}[hbtp]
\begin{center}
\includegraphics[width=0.49\textwidth]{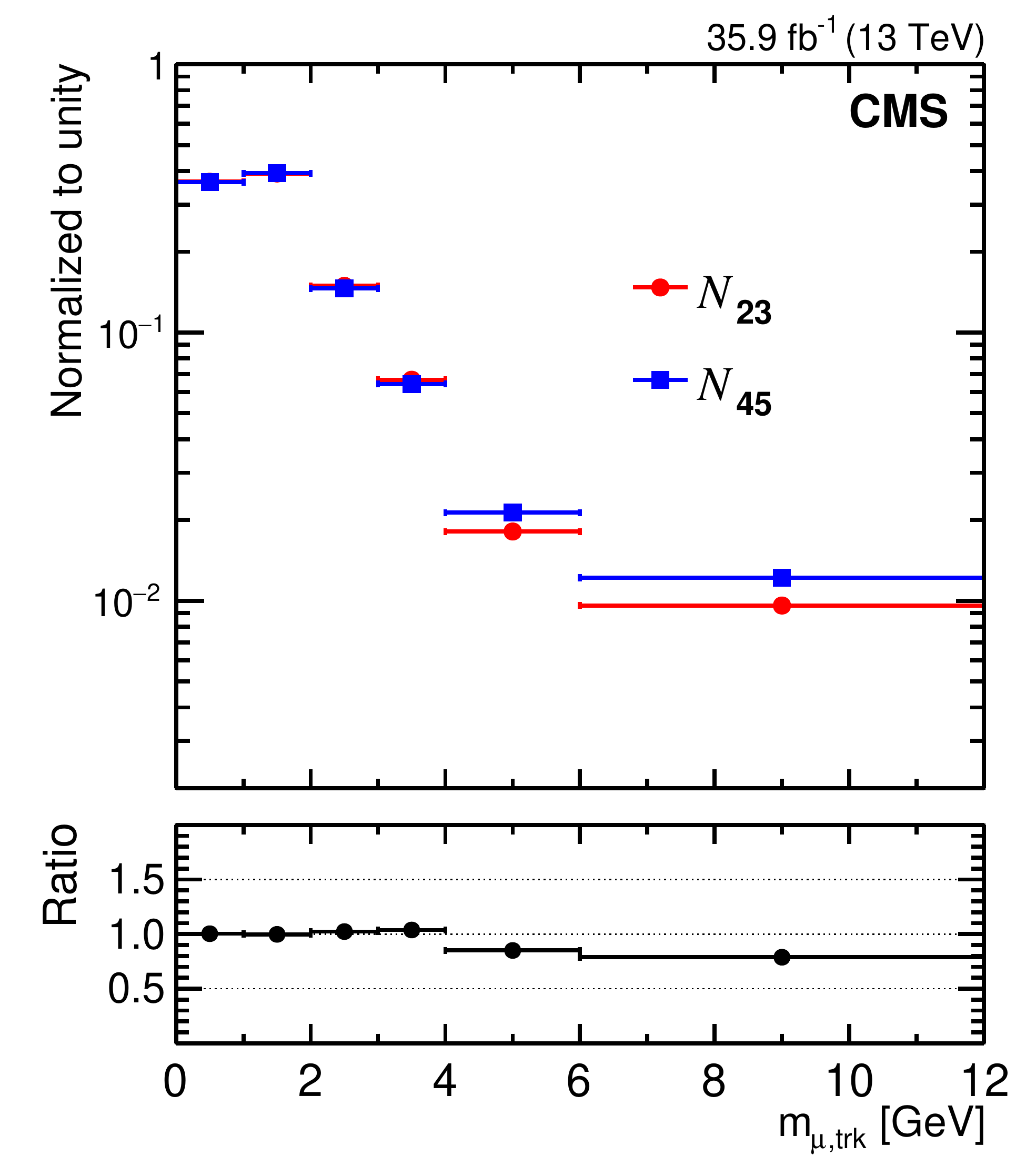}
\caption{
The observed invariant mass distribution, normalized to unity, of the muon-track invariant mass in control regions
$N_{23}$ (circles) and $N_{45}$ (squares).
}
\label{fig:n23_n45}
\end{center}
\end{figure}

The potential dependence of the muon-track invariant mass distribution
on the isolation requirement imposed on the second muon is verified also
by comparing shapes in the control regions $N_{23}$ and $N_{45}$.
The latter CR is defined by requiring the presence of 4 or 5 ``isolation" tracks
nearby to the second muon, while the first muon-track pair passes selection criteria
for the \PaO candidate. The results are illustrated in Fig.~\ref{fig:n23_n45}.
A slight difference is observed between distributions in these two CRs. This difference is taken as a shape uncertainty in
the normalized template $f_\text{1D}(j)$ entering Eq.~(\ref{eq:QCDshape}).

\begin{figure}[hbt!]
\begin{center}
\includegraphics[width=\cmsFigWidth]{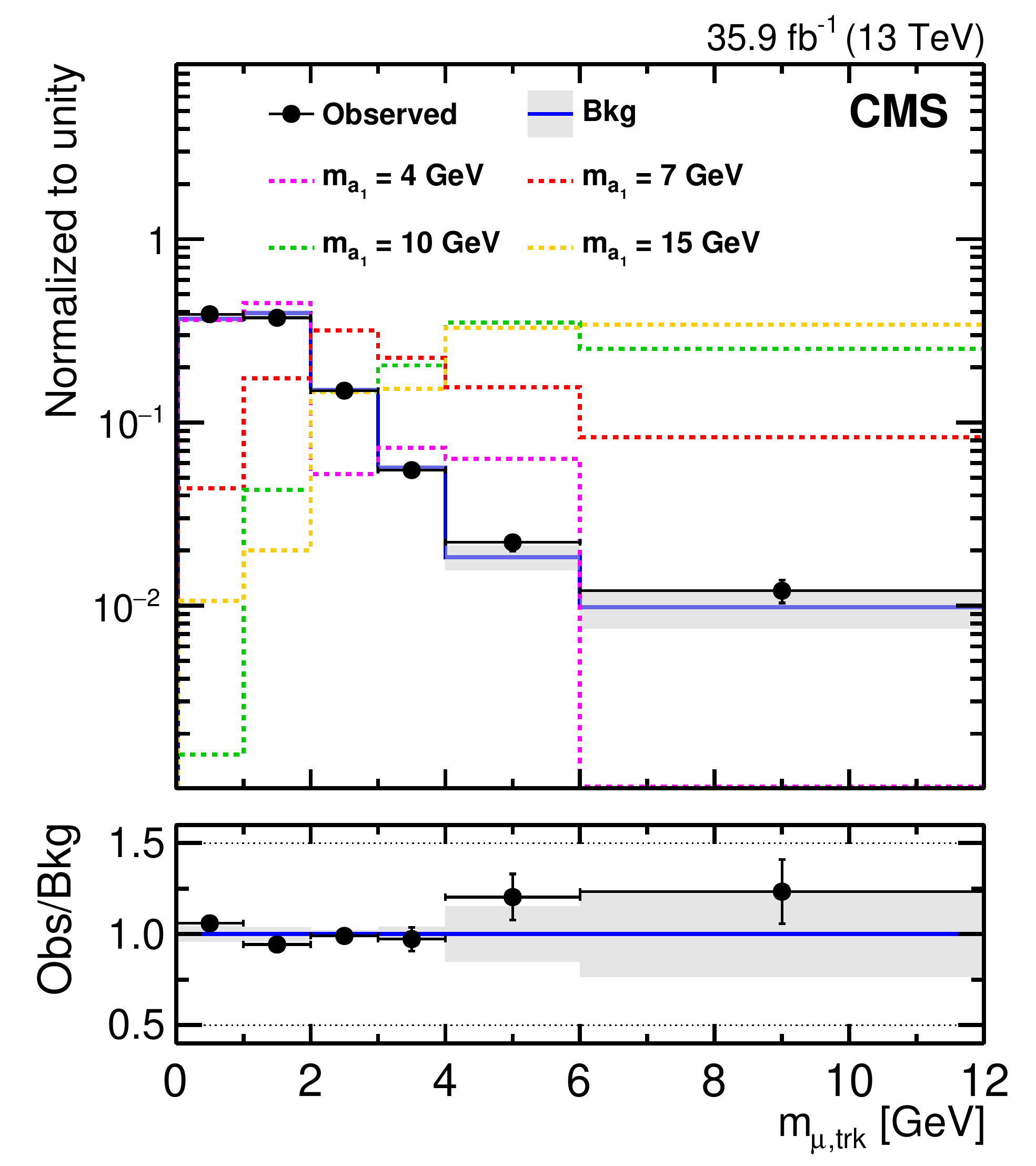}
\end{center}
\caption{
Normalized invariant mass distribution of the muon-track system for events passing the signal selection.
Observed numbers of events are represented by data points with error bars. The QCD multijet background model is derived from the
control region $N_{23}$. Also shown are the normalized distributions from signal simulations for
four mass hypotheses, $m_{\PaO}$= 4, 7, 10, and $15\GeV$ (dashed histograms), whereas for higher masses the analysis has no sensitivity.
Each event in the observed and expected signal distributions contributes two entries,
corresponding to the two muon-track systems in each event passing the selection. The signal distributions
include $2\mu2\tau$ and $4\tau$ contributions.
The lower panel shows the ratio of the observed to expected number of
background events in each bin of the distribution. The grey shaded area represents the background model uncertainty.}
\label{fig:MassShapes}
\end{figure}

Figure~\ref{fig:MassShapes} presents the normalized
invariant mass distribution of the muon-track system for
data selected in the SR and for the background model derived from the $N_{23}$ CR.
The data and background distributions are
compared to the signal distributions, obtained from simulation,
for four representative mass hypotheses, $m_{\PaO}$= 4, 7, 10, and 15\GeV.
The invariant mass of the muon-track system is found to have higher discrimination power
between the background and the signal at higher $m_{\PaO}$. For lower masses, the signal shape
becomes more background like, resulting in a reduction of discrimination power.

\subsection{Modeling of \texorpdfstring{$C(i,j)$}{C(i,j)}}
\label{Sec:bkdg_corr}

\begin{figure}[hbt!]
\begin{center}
\includegraphics[width=\cmsFigWidth]{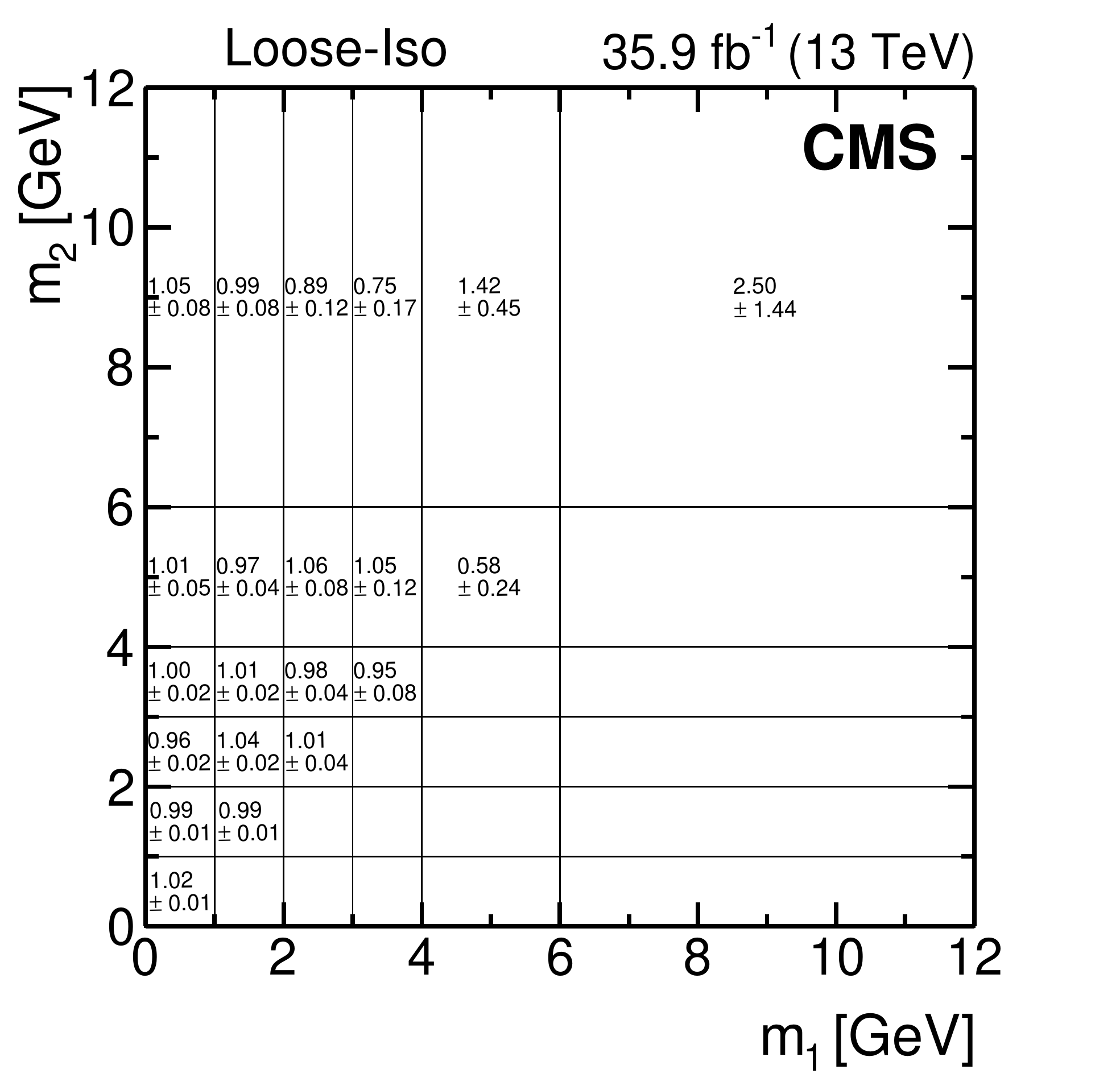}
\caption{
The ($m_1,m_2$) correlation factors $C(i,j)$ with their statistical uncertainties,
derived from data in the CR {{Loose-Iso}}.}
\label{fig:CorrCoeff}
\end{center}
\end{figure}

\begin{figure}[htb!]
\begin{center}
\includegraphics[width=0.49\textwidth]{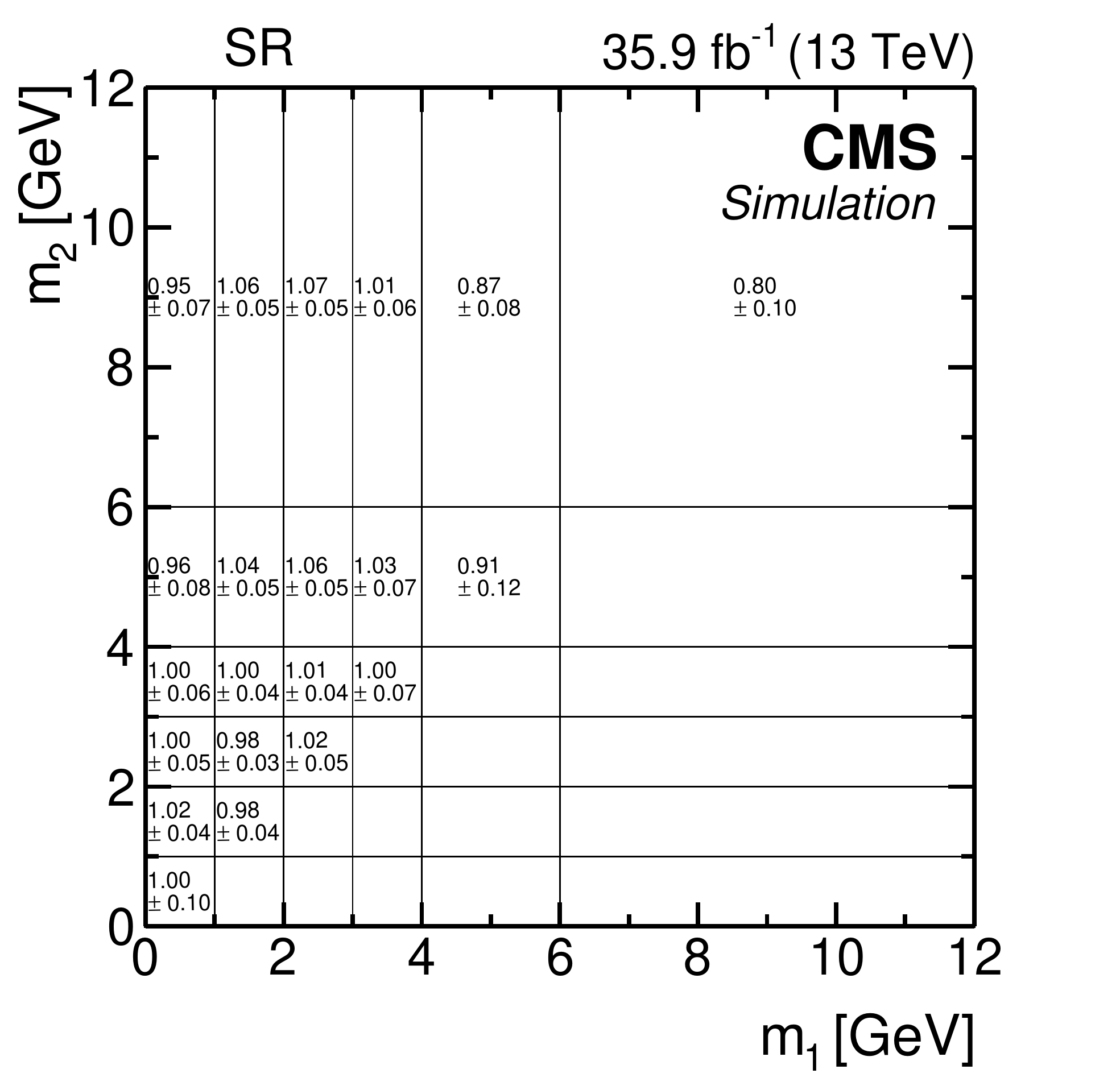}
\includegraphics[width=0.49\textwidth]{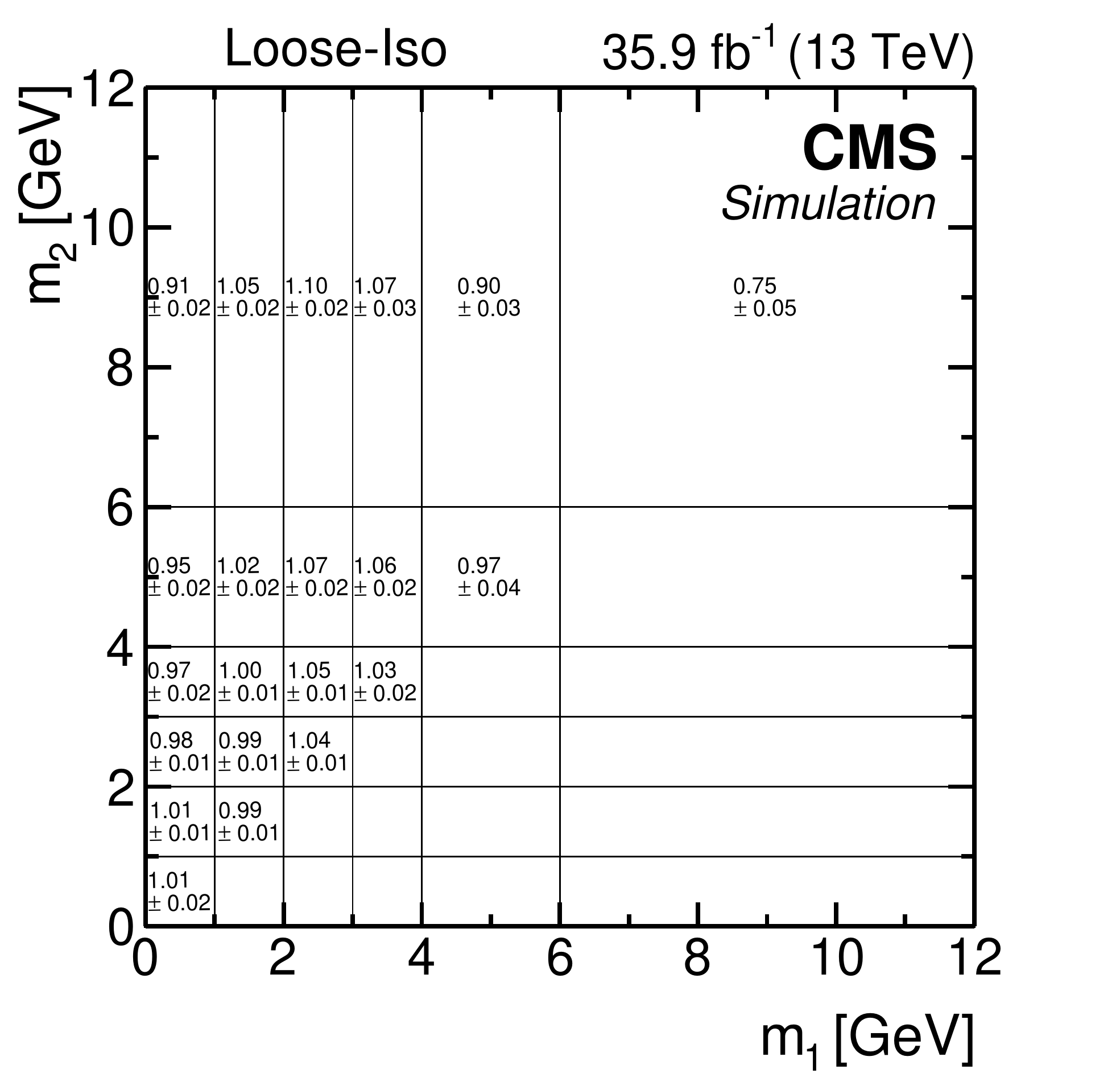}

\caption{
The ($m_1,m_2$) correlation factors $C(i,j)$ along with their MC statistical uncertainties,
derived from simulated samples in the (\cmsLeft: signal region,  \cmsRight: Loose-Iso CR).}
\label{fig:CorrCoeff_mc}
\end{center}
\end{figure}

In order to determine the correlation factors $C(i,j)$, an additional CR (labelled Loose-Iso)
is used. It consists of events that contain two SC muons
passing the identification and kinematic selection criteria outlined in Section~\ref{Sec:Selection}.
Each muon is required to have
two or three nearby tracks. One of them should belong to the category of ``signal" tracks, whereas
remaining tracks should belong to the category of ``soft" tracks.
About 36k data events are selected in this CR.
The simulation predicts that the QCD multijet events dominate this CR,
comprising more than 99\% of selected events. It was also found that the overall background-to-signal
ratio is enhanced compared to the SR by a factor of
30 to 40, depending on the mass hypothesis, $m_{\PaO}$. The event sample in this region is used to
build the normalized distribution $f_\text{2D}(i,j)$.
Finally, the correlation factors $C(i,j)$ are obtained according to Eq.~(\ref{eq:QCDshape}) as
\begin{equation}
C(i,j) = \frac{f_\text{2D}(i,j)}{(f_\text{1D}(i) f_\text{1D}(j))^{\text{sym}}},
\label{eq:Cij}
\end{equation}
where $f_\text{1D}(i)$ is the 1D normalized distribution with two entries per event ($m_1$ and $m_2$).
The correlation factors $C(i,j)$ derived from data in the Loose-Iso CR are presented in Fig.~\ref{fig:CorrCoeff}.
To obtain estimates of $C(i,j)$ in the signal region, the correlation factors derived in the Loose-Iso CR
have to be corrected for the difference in $C(i,j)$ between the signal region and Loose-Iso CR.
This difference is assessed by comparing samples of simulated background events.
The correlation factors estimated from simulation in the signal
region and the Loose-Iso CR  are presented in Fig.~\ref{fig:CorrCoeff_mc}.

The correlation factors in the signal region are then computed as
\begin{equation}
C(i,j)^{\mathrm{SR}}_{\mathrm{data}} = C(i,j)^{\mathrm{CR}}_{\mathrm{data}}\frac{C(i,j)^{\mathrm{SR}}_{\mathrm{MC}}}{C(i,j)^{\mathrm{CR}}_{\mathrm{MC}}},
\label{eq:Cij_extrap}
\end{equation}
where

\begin{itemize}
\item $C(i,j)^{\mathrm{CR}}_{\mathrm{data}}$ are correlation factors derived for the Loose-Iso CR in data (Fig.~\ref{fig:CorrCoeff});

\item $C(i,j)^{\mathrm{SR}}_{\mathrm{MC}}$ are correlation factors derived for the SR in the simulated QCD multijet sample (Fig.~\ref{fig:CorrCoeff_mc}, \cmsLeft);

\item $C(i,j)^{\mathrm{CR}}_{\mathrm{MC}}$ are correlation factors derived for the Loose-Iso CR in the simulated QCD multijet sample (Fig.~\ref{fig:CorrCoeff_mc}, \cmsRight).
\end{itemize}

The difference in correlation factors derived in the SR (Fig.~\ref{fig:CorrCoeff_mc}, \cmsLeft) and in the Loose-Iso CR
(Fig.~\ref{fig:CorrCoeff_mc}, \cmsRight) using the QCD multijet sample is taken into account as an uncertainty in $C(i,j)$.

\section{Modeling signal}

\label{Sec:Signalmodeling}

The signal templates are derived from the simulated samples of the $\PH\to\PaO\PaO\to 4\tau$
and $\PH\to\PaO\PaO\to 2\mu 2\tau$ decays.
The study probes the signal strength modifier,
defined as the ratio of the product of the measured signal cross section and the branching fraction
into the $4\tau$ final state
${\mathcal{B}}(\PH\to\PaO\PaO){\mathcal{B}}^{2}(\PaO\to \tau\tau)$ to the inclusive cross
section of the $\PH$  production predicted in the SM.
The relative contributions from different production modes of $\PH$
are defined by the corresponding cross sections predicted in the SM. The contribution
of the $\PH\to\PaO\PaO\to 2\mu 2\tau$ decay, is computed
assuming that the partial widths of $\PaO\to\tau\tau$ and $\PaO\to\mu\mu$ decays satisfy Eq.~(\ref{eq:width_ratios}).

\begin{figure}[hbt!]
\begin{center}
\includegraphics[width=0.49\textwidth]{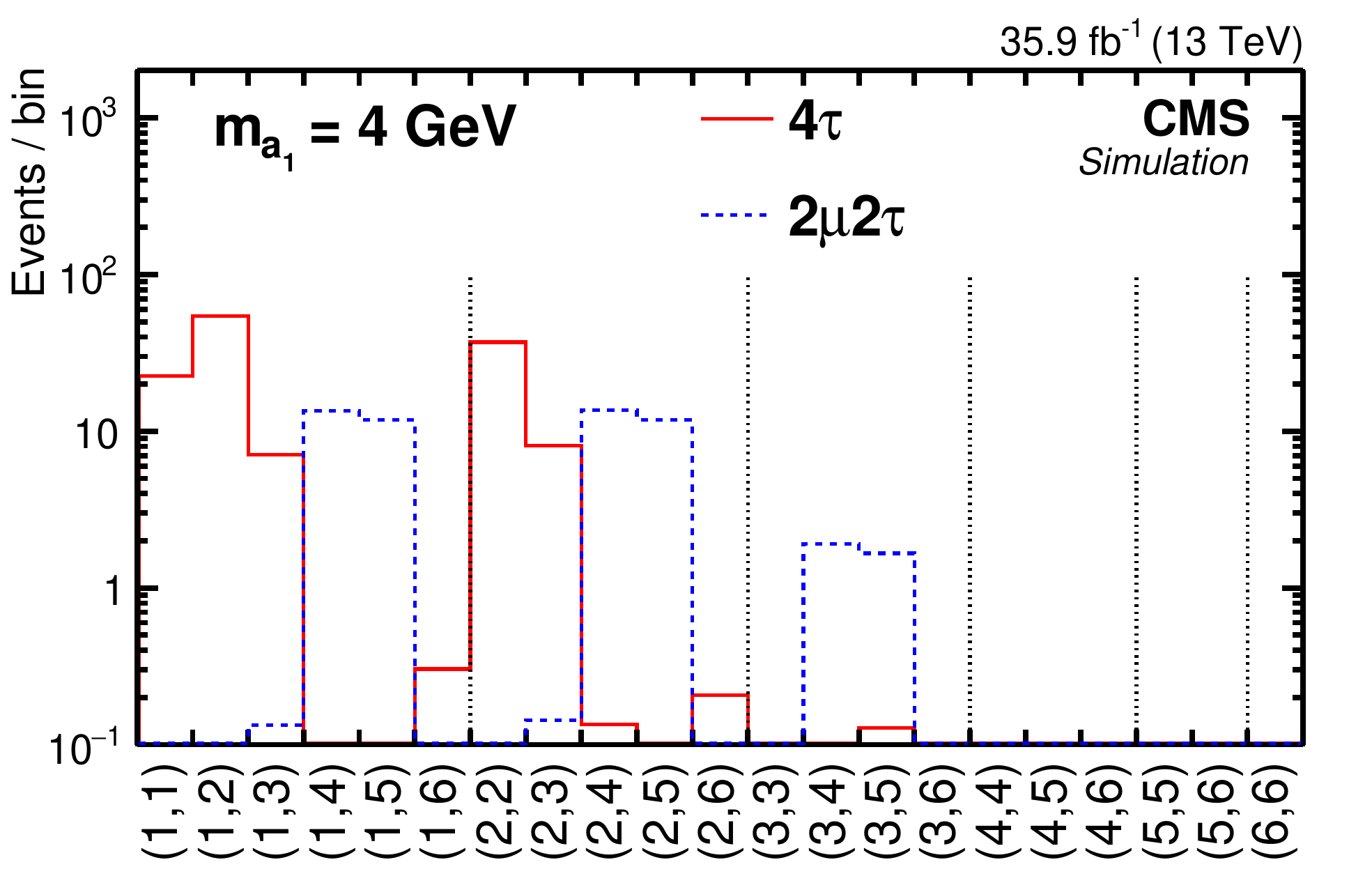}
\includegraphics[width=0.49\textwidth]{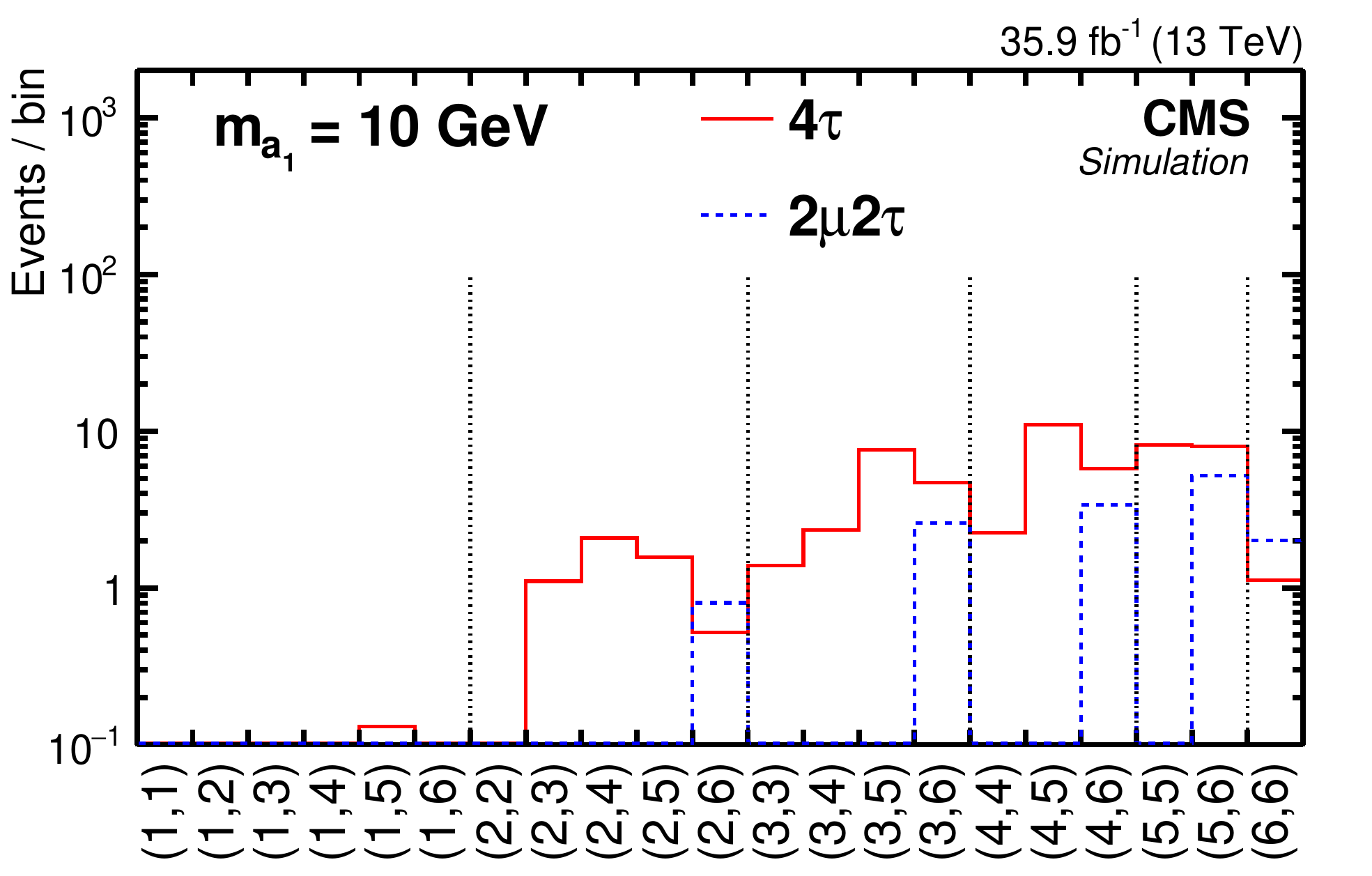}
\caption{
The distribution of the signal templates $f_\text{2D}(i,j)$ in one row for mass hypothesis $m_{\PaO}=4\GeV$ (\cmsLeft) and 10\GeV (\cmsRight).
The $\PH\to\PaO\PaO\to 2\mu 2\tau$ (blue histogram) and $\PH\to\PaO\PaO\to 4\tau$ (red histogram) contributions are shown. The notation
of the bins follows that of Fig.~\ref{fig:binning}.
}
\label{fig:signal_4tau_2mu2tau}
\end{center}
\end{figure}

The invariant mass distribution of the muon-track system in the $\PaO\to\mu\mu$ decay channel peaks at the nominal value of the \PaO boson mass,
while the reconstructed mass of the muon-track system in the $\PaO\to\tau\tau$ decay is typically lower, because of the missing neutrinos.
This is why the $\PH\to\PaO\PaO\to 2\mu 2\tau$ signal samples have a largely different shape of the ($m_1,m_2$)
distribution compared to the $\PH\to\PaO\PaO\to 4\tau$ signal samples.
Figure~\ref{fig:signal_4tau_2mu2tau} compares the ($m_1,m_2$) distributions unrolled in a one row between the $\PH\to\PaO\PaO\to 4\tau$ and
$\PH\to\PaO\PaO\to 2\mu 2\tau$ signal samples for mass hypotheses $m_{\PaO}4\GeV$ and 10\GeV.
The signal distributions are normalized assuming the SM $\PH$ production rate with the branching fraction ${\mathcal{B}} (\PH\to\PaO\PaO){\mathcal{B}}^{2}(\PaO\to \tau\tau)$ equal to 0.2.

\section{Systematic uncertainties}
\label{Sec:Systematics}

Table~\ref{tab:systematics} lists the systematic uncertainties considered in the analysis for both signal and background.

\begin{table*}[hbtp]
\topcaption{Systematic uncertainties and their effect on the estimates of the
QCD multijet background and signal.
}
\label{tab:systematics}
\begin{center}
\begin{tabular}{lcccc}
\hline
Source & Value & Affected & Type & Effect on the  \\
       &       & sample   &      & total yield    \\
\hline
Stat. unc. in $C(i,j)$ & 3--60\%     & bkg.   & bin-by-bin & \NA\\
Extrapolation unc. in $C(i,j)$ &  \NA        & bkg.   & shape      & \NA\\
Unc. in $f_\text{1D}(i)$  &  \NA        & bkg.   & shape      & \NA\\
Integrated luminosity     &  2.5\%      & signal & norm.      & 2.5\% \\
Muon id. and trigger efficiency        & 2\% per muon       & signal & norm. & 4\% \\
Track id. efficiency      & 4--12\% per track  & signal & shape & 10--18\%  \\
MC stat. unc. in signal yields & 8--100\%      & signal & bin-by-bin & 5--20\% \\
[\cmsTabSkip]
\multicolumn{5}{c}{Theoretical uncertainties in the signal acceptance} \\
$\mu_\text{R}$ and $\mu_\text{F}$ variations &  & signal & norm. & 0.8--2\% \\
PDF                                          &  & signal & norm. &   1--2\% \\
\multicolumn{5}{c}{Theoretical uncertainties in the signal cross sections} \\
$\mu_\text{R,F}$ variations (\PgPg)  & 5--7\%  & signal & norm. & 5--7\% \\
$\mu_\text{R,F}$ variations (other processes) & 0.4--9\%              & signal & norm. & $<$0.5\% \\
PDF (\PgPg)  &      3.1\%          & signal & norm. &    3.1\% \\
PDF (other processes) & 2.1--3.6\%          & signal & norm. & $<$0.5\% \\
\hline
\end{tabular}
\end{center}
\end{table*}

\subsection{Uncertainties related to the background}

The estimation of the QCD multijet background is based on observed data, therefore it is not
affected by imperfections in the simulation, reconstruction, or detector response.

The shape of the background in the
($m_1,m_2$) distribution is modeled according
to Eq.~(\ref{eq:QCDshape}), while its uncertainty is dominated by uncertainties related
to the correlation factors $C(i,j)$ (as described in Section~\ref{Sec:bkdg_corr}).
Additionally, it is also affected by the shape uncertainty in the
1D template $f_\text{1D}(m)$ (as discussed in Section~\ref{Sec:bkgd_1D}).
The bin-by-bin uncertainties in mass correlation factors $C(i,j)$, derived from Eq.~(\ref{eq:Cij_extrap}),
are composed of the statistical uncertainties in observed data and simulated samples, as presented in Figs.~\ref{fig:CorrCoeff} and~\ref{fig:CorrCoeff_mc}, and
range from 3 to 60\%. These uncertainties are accounted for in the
signal extraction procedure by one nuisance parameter per bin in the ($m_1,m_2$) distribution~\cite{Conway-PhyStat}.
The systematic uncertainties related to the extrapolation of $C(i,j)$ from the Loose-Iso CR
to the SR are derived from the dedicated MC study outlined in Section~\ref{Sec:bkdg_corr}.
The related shape uncertainty is determined by comparing correlation factors
derived in the simulated samples, between the signal region and the
Loose-Iso CR.

In the case when
\mbox{${\mathcal{B}}(\PH \to \PaO \PaO)$}${\mathcal{B}}^{2} (\PaO \to \tau \tau)=0.34$,
corresponding to an upper limit at 95\% confidence level (\CL)
on the branching fraction of the $\PH$ decay into non-SM particles from
Ref.~\cite{Khachatryan:2016vau}, the impact of possible signal
contamination in the {{Loose-Iso}} CR is estimated on a bin-by-bin basis,
and it is at most 2.8\% in the bin (6,6) which was found to have a negligible effect on the final results. 
For all other CRs, the signal contamination was found to be well below 1\%.

\subsection{Uncertainties related to signal}

An uncertainty of 2.5\% is assigned to the integrated luminosity estimate~\cite{CMS-PAS-LUM-17-001}.

The uncertainty in the muon identification and trigger
efficiency is estimated to be 2\% for each selected muon obtained with the tag-and-probe technique~\cite{CMS:2011aa}.
The track selection and muon-track isolation efficiency is assessed with a study
performed on a sample of \PZ bosons decaying into a pair of tau leptons.
In the selected $\PZ\to\tau\tau$ events, one tau lepton
is identified via its muonic decay, while the other is identified as an isolated track resulting
from a one-prong decay. The track is required to pass the nominal selection criteria used in the
main analysis. From this study, the uncertainty in the track selection and isolation efficiency
is evaluated. The related uncertainty affects the shape of the signal estimate, while changing the overall signal yield by 10--18\%.
The muon and track momentum scale uncertainties are smaller than 0.3\%
and have a negligible effect on the analysis.

The bin-by-bin statistical uncertainties in the signal acceptance range from 8 to 100\%,
while the impact on the overall signal normalization varies between 5 and 20\%.

Theoretical uncertainties have an impact on the differential kinematic distributions of the
produced $\PH$, in particular its \pt spectrum, thereby affecting signal acceptance.
The uncertainty due to missing higher-order corrections to the \PgPg
process is estimated with the \textsc{HqT} program by varying the renormalization ($\mu_\text{R}$) and
factorization ($\mu_\text{F}$) scales. The $\PH$ \pt-dependent $K$ factors are recomputed
according to these variations and applied to the simulated signal samples. The resulting
effect on the signal acceptance is estimated to vary between 1.2 and 1.5\%, depending on {$m_{\PaO}$}.
In a similar way, the uncertainty in the signal acceptance is computed for the VBF, VH and {$\ttbar\PH$}
production processes. The impact on the acceptance is estimated to vary between 0.8 and 2.0\%, depending on
the process and probed mass of the \PaO boson.

The \textsc{HqT} program is also used to evaluate the effect of the PDF uncertainties.
The nominal $K$ factors for the $\PH$  \pt spectrum
are computed with the NNPDF3.0 PDF set~\cite{Ball:2014uwa}.
Variations of the NNPDF3.0 PDFs within their uncertainties change the signal acceptance by about 1\%, whilst using the CTEQ6L1 PDF set~\cite{Pumplin:2002vw}
changes the signal acceptance by about 0.7\%. The impact of the PDF uncertainties on the acceptance for the VBF, VH and {$\ttbar\PH$}
production processes is estimated in the same way and a 2\% uncertainty is considered to account for these.

Systematic uncertainties in theoretical predictions for the signal cross sections are driven by
variations of the $\mu_\text{R}$ and $\mu_\text{F}$ scales and PDF uncertainties. Uncertainties
related to scale variations range from 0.4 to 9\%, depending on the production mode.
Uncertainties related to PDF vary between 2.1 and 3.6\%.

\section{Results}
\label{Sec:Results}

The signal is extracted with a binned maximum-likelihood fit applied to
the ($m_1,m_2$) distribution. For each probed
mass of the \PaO boson, the ($m_1,m_2$) distribution is fitted with the sum of two templates, corresponding
to expectations for the signal and background, dominated by QCD multijet events.

The normalization of both signal and background are allowed to float freely in the fit. The
systematic uncertainties affecting the normalization of the signal templates are incorporated in the
fit via nuisance parameters with a log-normal prior probability density function. The shape-altering
systematic uncertainties are represented by nuisance parameters whose variations cause
continuous morphing of the signal or background template shape, and are assigned a Gaussian prior probability
density functions. The bin-by-bin statistical uncertainties are assigned gamma prior probability
density functions.

\begin{figure*}[hbt!]
\begin{center}{
\includegraphics[width=0.7\textwidth]{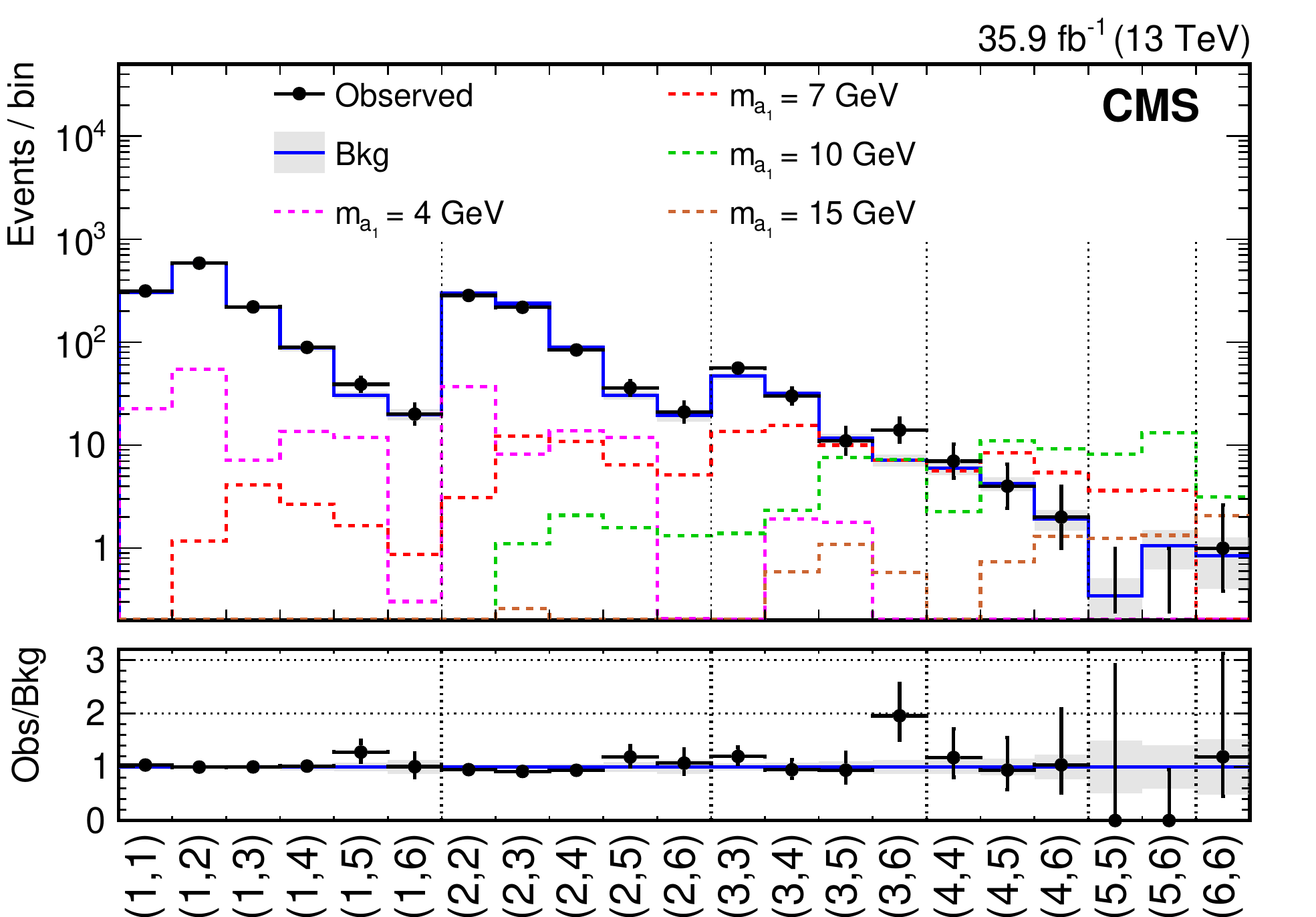}
\caption{
The ($m_1,m_2$) in one row distribution used to extract the signal.
Observed numbers of events are represented by data points with error bars. The background with its uncertainty is shown as the blue histogram with the shaded
error band. The shape and the normalization of the background distribution are
obtained by applying a fit to the observed data under the background-only hypothesis.
Signal expectations for the $4 \tau$ and $2\mu2\tau$ final states are shown as dotted
histograms for the mass hypotheses $m_{\PaO}$= 4, 7, 10 and 15\GeV. The relative normalization of the $4 \tau$ and $2\mu2\tau$ final states are given by Eq.~(\ref{eq:width_ratios}) as explained in Section~\ref{Sec:Signalmodeling}. The signal normalization is computed assuming that the
$\PH$ boson is produced in $\Pp\Pp$ collisions with a rate predicted by the SM, and decays
into $\PaO \PaO \to 4\tau$ final state with the branching fraction of 20\%. The lower plot shows the ratio of the observed data events to
the expected background yield in each bin of the ($m_1,m_2$) distribution.
}
\label{fig:AnalysisBins}
}
\end{center}
\end{figure*}

Figure~\ref{fig:AnalysisBins} shows the distribution of ($m_1,m_2$), where the
notation for the bins follows that of Fig.~\ref{fig:binning}. The shape and the normalization of the background distribution
are obtained by applying a fit to the observed data under the background-only hypothesis. Also shown are the expectations for
the signal at $m_{\PaO}$= 4, 7, 10, and 15\GeV. The signal normalization is computed assuming that
the $\PH$  is produced in $\Pp\Pp$ collisions with a rate predicted by the standard model, and decays
into $\PaO \PaO \to 4\tau$ final state with a branching fraction of 20\%. No significant deviations from the background expectation
are observed in the ($m_1,m_2$) distribution.

\begin{figure}[hbt!]
\begin{center}{
\includegraphics[width=\cmsFigWidth]{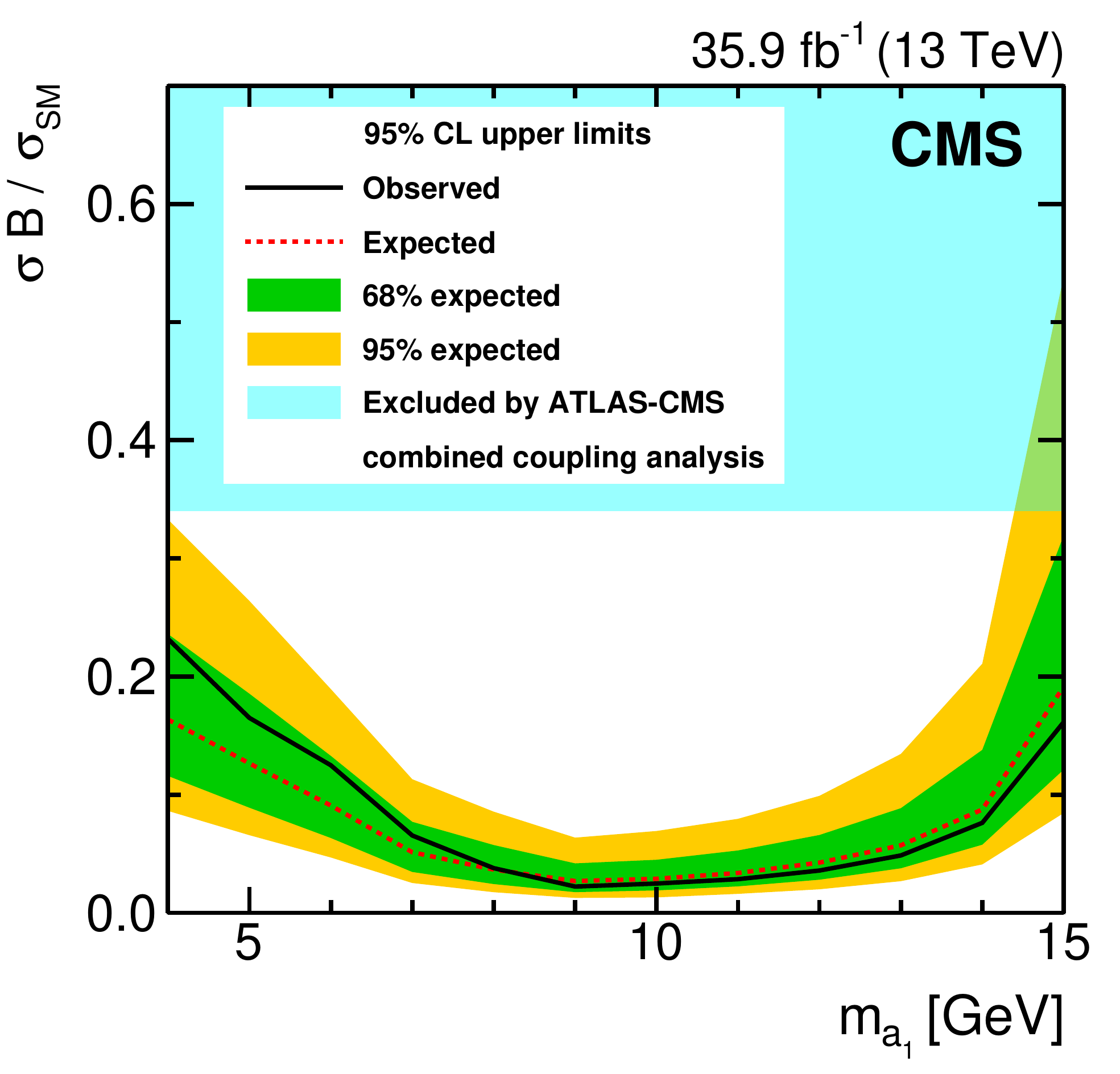}
\caption{
The observed and expected upper limits at 95\% confidence levels on the product of signal cross section and the branching fraction
$\sigma (\Pp\Pp \to \PH+X)   {\mathcal{B}} (\PH \to \PaO \PaO)   {\mathcal{B}}^{2} (\PaO \to \tau \tau)$,
relative to the inclusive Higgs boson production cross section $\sigma_\text{SM}$ predicted in the SM. The green and yellow bands indicate the regions that contain 68\% and 95\% of the distribution of limits expected under the background-only hypothesis. The shaded area in blue
indicates the excluded region of $>$34\% for the branching fraction of the $\PH$ decay into non-SM particles at 95\% \CL from Ref.~\cite{Khachatryan:2016vau}.
\label{fig:limits}
}
}
\end{center}
\end{figure}

Results of the analysis are used to set upper limits at 95\% \CL on the product of the cross section and branching fraction,
$\sigma (\Pp\Pp \to \PH+X)   {\mathcal{B}} (\PH \to \PaO \PaO)   {\mathcal{B}}^{2} (\PaO \to \tau \tau)$,
relative to the inclusive SM cross section of $\PH$ production.
The modified frequentist \CLs criterion~\cite{Read:2002hq,Junk:1999kv}, and
the asymptotic formulae are used for the test statistic~\cite{Cowan:2010js}, implemented in
the {\textsc RooStats} package~\cite{Moneta:2010pm}. Figure~\ref{fig:limits} shows the observed and expected upper limits at 95\% \CL on the signal
cross section times the branching fraction, relative to the total cross section of the
$\PH$ boson production as predicted in the SM.
The observed limit is compatible with the expected limit within one standard deviation
in the entire range of $m_{\PaO}$ considered, and ranges from 0.022 at $m_{\PaO}=9$\GeV to 0.23 at $m_{\PaO}=4$\GeV and reaches 0.16 at $m_{\PaO}=15$\GeV.
The expected upper limit ranges from 0.027 at $m_{\PaO}=9$\GeV to 0.16 at $m_{\PaO}=4$\GeV and reaches 0.19 at $m_{\PaO}=15$\GeV. The degradation of the analysis sensitivity towards lower values of $m_{\PaO}$ is caused
by the increase of the background yield at low invariant masses of the muon-track systems, as illustrated in Figs.~\ref{fig:MassShapes} and~\ref{fig:AnalysisBins}.
With increasing $m_{\PaO}$, the average angular separation between the decay products of the \PaO boson is increasing. As a consequence, the efficiency of the
signal selection drops down, as we require the muon and the track, originating from the $\PaO\to\tau_\mu\tau_{\text{one-prong}}$ or $\PaO\to\mu\mu$ decay,
to be within a cone of $\Delta{\mathrm{R}}=0.5$. This explains the deterioration of the search sensitivity at higher values of $m_{\PaO}$.  The shaded area in blue
indicates the excluded region of $>$34\% for the branching fraction of the $\PH$ decay into non-SM particles at 95\% \CL~\cite{Khachatryan:2016vau}.

The new limits improve significantly over the previous 8\TeV limits~\cite{Khachatryan:2015nba} by 30\% (for low masses) and up to 80\% (for intermediate masses of 8\GeV), while the new analysis further extends the coverage of $m_{\PaO}$ up to 15\GeV.

\section{Summary}
A search is presented for light pseudoscalar \PaO bosons, produced from decays of the 125\GeV Higgs boson ($\PH$) in a data set corresponding
to an integrated luminosity of 35.9\fbinv
of proton-proton collisions at a center-of-mass energy of 13\TeV.
The analysis is based on the $\PH$  inclusive production and targets
the $\PH\to \PaO \PaO \to 4\tau/2\mu 2\tau$ decay channels.
Both channels are used in combination to constrain
the product of the inclusive signal production cross section and the branching fraction into the $4\tau$ final state, exploiting
the linear dependence of the fermionic coupling strength of \PaO on the fermion mass.
With no evidence for a signal, the observed 95\% confidence level upper limit on the product of the inclusive
signal cross section and the branching fraction, relative to the SM $\PH$ production cross
section, ranges from 0.022 at $m_{\PaO}=9$\GeV to 0.23 at $m_{\PaO}=4$\GeV and
reaches 0.16 at $m_{\PaO}=15$\GeV. The expected upper limit ranges from 0.027 at $m_{\PaO}=9$\GeV
to 0.16 at $m_{\PaO}=4$\GeV and reaches 0.19 at $m_{\PaO}=15$\GeV.

\begin{acknowledgments}
We congratulate our colleagues in the CERN accelerator departments for the excellent performance of the LHC and thank the technical and administrative staffs at CERN and at other CMS institutes for their contributions to the success of the CMS effort. In addition, we gratefully acknowledge the computing centers and personnel of the Worldwide LHC Computing Grid for delivering so effectively the computing infrastructure essential to our analyses. Finally, we acknowledge the enduring support for the construction and operation of the LHC and the CMS detector provided by the following funding agencies: BMBWF and FWF (Austria); FNRS and FWO (Belgium); CNPq, CAPES, FAPERJ, FAPERGS, and FAPESP (Brazil); MES (Bulgaria); CERN; CAS, MoST, and NSFC (China); COLCIENCIAS (Colombia); MSES and CSF (Croatia); RPF (Cyprus); SENESCYT (Ecuador); MoER, ERC IUT, PUT and ERDF (Estonia); Academy of Finland, MEC, and HIP (Finland); CEA and CNRS/IN2P3 (France); BMBF, DFG, and HGF (Germany); GSRT (Greece); NKFIA (Hungary); DAE and DST (India); IPM (Iran); SFI (Ireland); INFN (Italy); MSIP and NRF (Republic of Korea); MES (Latvia); LAS (Lithuania); MOE and UM (Malaysia); BUAP, CINVESTAV, CONACYT, LNS, SEP, and UASLP-FAI (Mexico); MOS (Montenegro); MBIE (New Zealand); PAEC (Pakistan); MSHE and NSC (Poland); FCT (Portugal); JINR (Dubna); MON, RosAtom, RAS, RFBR, and NRC KI (Russia); MESTD (Serbia); SEIDI, CPAN, PCTI, and FEDER (Spain); MOSTR (Sri Lanka); Swiss Funding Agencies (Switzerland); MST (Taipei); ThEPCenter, IPST, STAR, and NSTDA (Thailand); TUBITAK and TAEK (Turkey); NASU and SFFR (Ukraine); STFC (United Kingdom); DOE and NSF (USA).

\hyphenation{Rachada-pisek} Individuals have received support from the Marie-Curie program and the European Research Council and Horizon 2020 Grant, contract Nos.\ 675440, 752730, and 765710 (European Union); the Leventis Foundation; the A.P.\ Sloan Foundation; the Alexander von Humboldt Foundation; the Belgian Federal Science Policy Office; the Fonds pour la Formation \`a la Recherche dans l'Industrie et dans l'Agriculture (FRIA-Belgium); the Agentschap voor Innovatie door Wetenschap en Technologie (IWT-Belgium); the F.R.S.-FNRS and FWO (Belgium) under the ``Excellence of Science -- EOS" -- be.h project n.\ 30820817; the Beijing Municipal Science \& Technology Commission, No. Z181100004218003; the Ministry of Education, Youth and Sports (MEYS) of the Czech Republic; the Lend\"ulet (``Momentum") Program and the J\'anos Bolyai Research Scholarship of the Hungarian Academy of Sciences, the New National Excellence Program \'UNKP, the NKFIA research grants 123842, 123959, 124845, 124850, 125105, 128713, 128786, and 129058 (Hungary); the Council of Science and Industrial Research, India; the HOMING PLUS program of the Foundation for Polish Science, cofinanced from European Union, Regional Development Fund, the Mobility Plus program of the Ministry of Science and Higher Education, the National Science Center (Poland), contracts Harmonia 2014/14/M/ST2/00428, Opus 2014/13/B/ST2/02543, 2014/15/B/ST2/03998, and 2015/19/B/ST2/02861, Sonata-bis 2012/07/E/ST2/01406; the National Priorities Research Program by Qatar National Research Fund; the Ministry of Science and Education, grant no. 3.2989.2017 (Russia); the Programa Estatal de Fomento de la Investigaci{\'o}n Cient{\'i}fica y T{\'e}cnica de Excelencia Mar\'{\i}a de Maeztu, grant MDM-2015-0509 and the Programa Severo Ochoa del Principado de Asturias; the Thalis and Aristeia programs cofinanced by EU-ESF and the Greek NSRF; the Rachadapisek Sompot Fund for Postdoctoral Fellowship, Chulalongkorn University and the Chulalongkorn Academic into Its 2nd Century Project Advancement Project (Thailand); the Welch Foundation, contract C-1845; and the Weston Havens Foundation (USA).
\end{acknowledgments}

\bibliography{auto_generated}

\cleardoublepage \appendix\section{The CMS Collaboration \label{app:collab}}\begin{sloppypar}\hyphenpenalty=5000\widowpenalty=500\clubpenalty=5000\input{HIG-18-006-authorlist.tex}\end{sloppypar}
\end{document}

%% file: HIG-18-006-authorlist.tex
\vskip\cmsinstskip
\textbf{Yerevan Physics Institute, Yerevan, Armenia}\\*[0pt]
A.M.~Sirunyan, A.~Tumasyan
\vskip\cmsinstskip
\textbf{Institut f\"{u}r Hochenergiephysik, Wien, Austria}\\*[0pt]
W.~Adam, F.~Ambrogi, E.~Asilar, T.~Bergauer, J.~Brandstetter, M.~Dragicevic, J.~Er\"{o}, A.~Escalante~Del~Valle, M.~Flechl, R.~Fr\"{u}hwirth\cmsAuthorMark{1}, V.M.~Ghete, J.~Hrubec, M.~Jeitler\cmsAuthorMark{1}, N.~Krammer, I.~Kr\"{a}tschmer, D.~Liko, T.~Madlener, I.~Mikulec, N.~Rad, H.~Rohringer, J.~Schieck\cmsAuthorMark{1}, R.~Sch\"{o}fbeck, M.~Spanring, D.~Spitzbart, W.~Waltenberger, J.~Wittmann, C.-E.~Wulz\cmsAuthorMark{1}, M.~Zarucki
% ...
\vskip\cmsinstskip
% ...
1:  Also at Vienna University of Technology, Vienna, Austria\\
% ...
